\documentclass[twocolumn]{aastex631}
\usepackage{graphicx}	
\usepackage{amsmath}	
\usepackage{amssymb}	
\usepackage[]{natbib}
\DeclareUnicodeCharacter{0301}{\hspace{-1ex}\'{ }}

\newcommand{\hii}{H\,{\scshape ii}}

\shorttitle{Star formation around Trumpler 37}
\shortauthors{Das et al.}

\begin{document}

\title{Membership analysis and 3D kinematics of the star-forming complex around Trumpler 37 using Gaia-DR3 \footnote{swagat@das.uchile.cl / dasswagat77@gmail.com}}

\author[0000-0001-7151-0882]{Swagat R. Das}
\affiliation{Departamento de Astronomı́a, Universidad de Chile, Las Condes, \\
7591245 Santiago, Chile}
\affiliation{Indian Institute of Science Education and Research (IISER) Tirupati, \\
Rami Reddy Nagar, Karakambadi Road, Mangalam (P.O.), \\
Tirupati 517507, India}
\author{Saumya Gupta}
\affiliation{Indian Institute of Science Education and Research (IISER) Tirupati, \\
Rami Reddy Nagar, Karakambadi Road, Mangalam (P.O.), \\
Tirupati 517507, India}
\author{Prem Prakash}
\affiliation{Department of Physics, Indian Institute of Technology (IIT) Hyderabad, India}
\author{Manash Samal}
\affiliation{Physical Research Laboratory, Ahmedabad, Gujrat, India}
\author{Jessy Jose}
\affiliation{Indian Institute of Science Education and Research (IISER) Tirupati, \\
Rami Reddy Nagar, Karakambadi Road, Mangalam (P.O.), \\
Tirupati 517507, India}
%



\begin{abstract}
Identifying and characterizing young populations of star-forming regions is crucial to
unravel their properties. In this regard, Gaia-DR3 data and machine learning tools are 
very useful for studying large star-forming complexes. In this work, we analyze the $\rm \sim7.1degree^2$ area
of one of our Galaxy's dominant feedback-driven star-forming complexes, i.e., the region around Trumpler 37. Using the Gaussian mixture and random forest classifier methods, we identify 1243 high-probable members in the complex, of which $\sim60\%$ are new members and are complete down to the mass limit of $\sim$0.1 $-$ 0.2~$\rm M_{\odot}$. The spatial distribution of the stars reveals multiple clusters towards the complex, where the central cluster around the massive star HD 206267 reveals two sub-clusters. Of the 1243 stars, 152 have radial velocity, with a mean value of $\rm -16.41\pm0.72~km/s$. We investigate stars' internal and relative movement within the central cluster. The kinematic analysis shows that the cluster's expansion is relatively slow compared to the whole complex. This slow expansion is possibly due to newly formed young stars within the cluster. We discuss these results in the context of hierarchical collapse and feedback-induced collapse mode of star formation in the complex.
\end{abstract}



\section{Introduction}\label{intro}
\begin{figure*}
\centering
\includegraphics[scale=0.5]{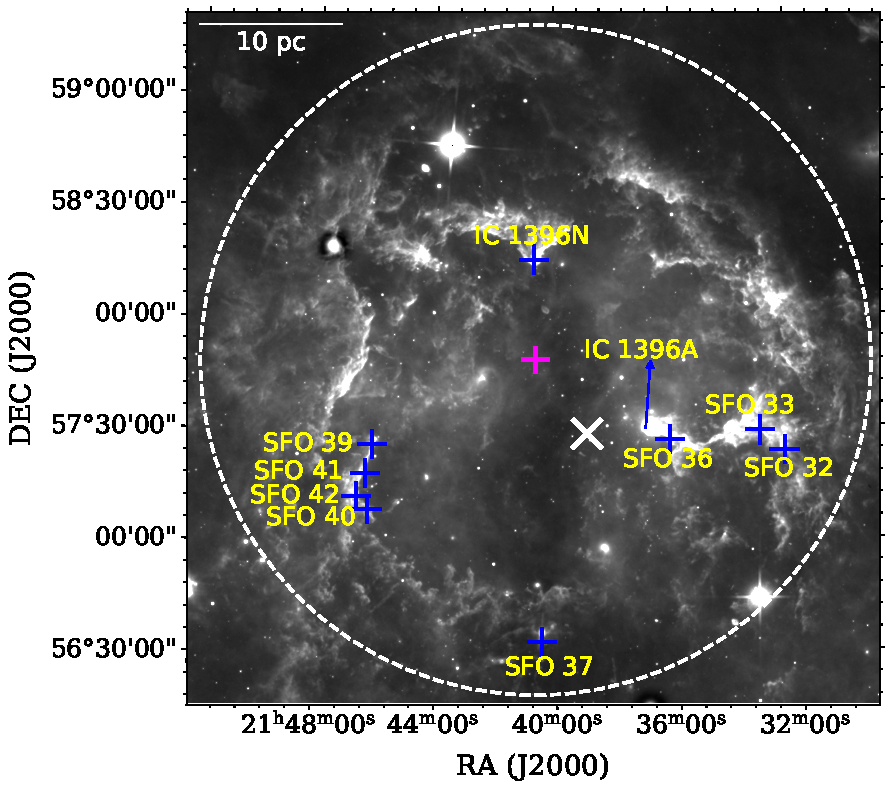}
\caption{This figure shows the WISE $\rm 22~\mu m$ image of the IC 1396. The cross mark (`$\times$') shows the position of the massive star HD 206267. The major globules from \citet{1991ApJS...77...59S} are marked as `+' symbols, and their IDs are mentioned. The white dashed circle (radius of $1.5^\circ$) is the region considered in this study for searching members using the Gaia-DR3 data. Center of the circle ($\rm \alpha = 21:40:39.28$ and $\rm \delta = +57:49:15.51$) marked as a magenta `+'. A scale bar of 10~pc is shown in the top-left corner.} 
\label{wise}
\end{figure*}

Star formation is one of the most complicated yet least understood phenomena in the field of astrophysics. Most of the stars form in clusters \citep{1964ARA&A...2..213B,1983MNRAS.203.1011E,1987IAUS..115....1L,2000prpl.conf..151C,
2004ApJS..154..367M,2008MNRAS.389.1556B,2010ARA&A..48..431P,2011MNRAS.410L...6G,2012MNRAS.419.2606B} by the fragmentation and hierarchical collapsing of molecular clouds \citep{1981MNRAS.194..809L,2004ARA&A..42..211E,2004RvMP...76..125M,2007ARA&A..45..565M}. 
Star clusters are unique tracers of galactic properties such as their origin, dynamics, and evolution \citep{2008IAUS..246...13K,2016ApJ...828...75F}. 
In addition to this, such studies aid in investigating the kinematics, dispersion, and evolution of the star-forming environment \citep{2019ApJ...870...32K,2019ApJ...871...46K,2020ApJ...900L...4P}. Clusters with massive O and B type stars serve as important laboratories for star-formation since these massive stars ionize their surroundings, create \hii\ regions and shape the evolution of low mass star population in the vicinity through their feedback effects \citep{2016ApJ...822...49J,2014A&A...566A.122S,2017MNRAS.472.4750D,2021MNRAS.500.3123D,2020A&A...638A...7Z,2021MNRAS.508.3388G,2022ApJ...926...25P}.
Hence, the identification and characterization of cluster members are essential to investigate various star-formation properties, such as stars form hierarchically by the natural collapse of clumpy molecular clouds or by the collapsing gas formed through sweeping and compression of the cold neutral gas by the \hii\ regions and bubbles. The distinction between these processes is important in understanding the net outcome of star formation, such as star formation efficiency (SFE) and star formation rate (SFR) due to various modes of star formation processes \citep{2012MNRAS.424..377D,2013MNRAS.430..234D,2015MNRAS.454..238W}.

The Global Astrometric Interferometer for Astrophysics (Gaia; \citealt{2016A&A...595A...1G}) data has revolutionized the identification and investigation of various scientific properties of the Galactic clusters \citep{2017MNRAS.470.2702K,2018A&A...616A..12G,
2019A&A...623A.108B,2019ApJ...870...32K,
2021MNRAS.504.2557D}. The Gaia-DR2 \citep{2018A&A...616A...1G} data contains the five parameters (positions, parallax, and proper motions) and astrometric solutions of $\sim$1.3 billion of stars up to G-band magnitude of 21 \citep{2018A&A...616A...1G}. 
Compared to Gaia-DR2, the Gaia-EDR3 improved the accuracy in proper motion and parallax measurements by factors of 2 and 2.5, respectively \citep{2020arXiv201201533G}. This accuracy improvement has benefited a better distinction of cluster members, especially for distant clusters. The final data release, Gaia-DR3, has significantly improved the radial velocity measurement of stars. The Gaia-DR3 preserves the astrometry properties of Gaia-EDR3, but it has improved the radial velocity measurement compared to the Gaia-DR2 in terms of accuracy and number of stars. This work aims to identify the new member population associated with the star-forming complex around Trumpler 37 (Tr 37) in IC 1396 using the multi-dimensional Gaia-DR3 data and machine learning techniques.

This work is arranged as follows. We describe the complex IC 1396 in Section \ref{source_detl}. In Section \ref{analy}, we present the analysis and results of this work. This includes the details of Gaia-DR3 data, the membership analysis using the machine learning approach, and the properties of the identified members. In Section \ref{res}, we discuss the various physical properties of IC 1396 derived using new members identified in this work along with literature-based members. We discuss the complex's 3D kinematic property and star-formation history in Section \ref{diss}. We summarize our work in Section \ref{summ}.
 \\ \\

\section{IC 1396} \label{source_detl}
The star-forming complex around Trumpler 37 (Tr 37; \citealt{1930LicOB..14..154T}) in IC 1396, shown in Figure \ref{wise}, is one of the classic examples of \hii\ regions with simple circular morphology, which is a part of the Cepheus OB2 complex \citep{1999AJ....117..354D}. IC 1396 has relatively low ($\rm A_V<$5~mag) foreground reddening \citep{2005AJ....130..188S,2012MNRAS.426.2917G,2012AJ....143...61N}. The star-forming complex is believed to be powered by the massive star (HD 206267) of spectral type O6 V, located near the center \citep{1995Obs...115..180S}. This \hii\ region is well known for its association with more than 20 bright-rimmed clouds (BRCs; \citealt{1991ApJS...77...59S}), fingertip structures, and elephant trunk structures in and around them, suggesting feedback effect from the massive central star \citep{1991ApJ...370..263S, 2005A&A...432..575F, 2012MNRAS.421.3206S}. The well-known BRCs at the peripheries of the \hii\ region (IC 1396A and IC 1396N) have often been referred to as the best examples of feedback-driven star formation \citep{2004AJ....128..805S,2006ApJ...638..897S,2007ApJ...654..316G,2010ApJ...717.1067C,
2013A&A...559A...3S,2014MNRAS.443.1614P,2014A&A...562A.131S,2019A&A...622A.118S}, with many previous studies focused around IC 1396A.
Using Gaia-DR2 data of the previously identified members, \citet{2019A&A...622A.118S} estimate a distance of $\rm 945^{+90}_{-73}~pc$, which is consistent within errors with the previous estimate of \citet{2002AJ....124.1585C}. Also, \citet{2005AJ....130..188S} obtained a mean age of $\rm \sim2 - 4~Myr$ of the complex based on the spectroscopically identified members. 
The modest distance and low foreground reddening make IC 1396 an ideal target for understanding the evolution of the \hii\ region and exploring the low-mass population associated with the complex.

We present the entire field of view of IC 1396 using the WISE $\rm 22~\mu m$ image in Figure \ref{wise}. The region exhibits a prominent mid-infrared cavity of radius $\sim$ $1.5^\circ$, which signifies the role of UV photons from the associated massive stars towards the gas and dust content of the cluster.
BRCs, fingertip, and elephant trunk structures are visible towards the periphery of the \hii\ region displaying the feedback-driven activity in the region. To better understand the evolution of the host \hii\ region and its possible impact on the next generation stars associated with BRCs/globules and hence the star formation history of the complex, it is important to identify the total member population of the whole complex.
There have been many studies in the past in search of the young stellar objects (YSOs) associated with the complex, however these surveys have different area coverage and sensitivity. A brief detail of the membership analysis from previous works towards the complex is given in the next subsection.

Gaia-DR3, due to its improvement in both photometry, astrometry, and radial velocity measurements over Gaia-DR2, is the best data set to obtain the membership population of the complex and, subsequently, its physical properties.

\subsection{Member population from previous studies} \label{lit_stars}

\begin{table*}
\small
\centering
\caption{Area covered and the number of stars obtained in previous works of literature.}
\label{work_mem}
\begin{tabular}{ccccc}
\\ \hline \hline
Work & No. of stars & Radius (degree) & RA (J2000) & DEC (J2000) \\
\hline
\citet{2002AJ....124.1585C}   & 66  & 0.5  & 21:39:09.89 & +57:30:56.07 \\
\citet{2006AJ....132.2135S}   & 172 & 0.6  & 21:37:54.41 & +57:33:15.32 \\
Sicilia-Aguilar et al. (2013) & 67  & 0.25 & 21:37:03.17 & +57:29:05.43 \\

\citet{2004ApJS..154..385R}   & 17  & 0.12  & 21:36:33.09 & +57:29:13.83 \\
\citet{2006ApJ...638..897S}   & 57  & 0.15 & 21:36:39.73 & +57:29:28.45 \\
\citet{2009ApJ...702.1507M}   & 69  & 0.15 & 21:36:36.32 & +57:29:54.78 \\

\citet{2011MNRAS.415..103B}   & 158 & 1.5 & 21:40:00.43 & +57:26:42.60 \\
\citet{2012AJ....143...61N}   & 639 & 1.4 & 21:39:48.76 & +57:30:31.56 \\

\citet{2007ApJ...654..316G}   & 24  & 0.1  & 21:40:36.73 & +58:15:37.51 \\
\citet{2009AJ....138....7M}   & 39  & 0.15 & 21:38:54.67 & +57:29:17.61 \\
\citet{2012MNRAS.426.2917G}   & 457 & 0.25 & 21:37:05.85 & +57:32:30.06  \\

\citet{2021AJ....162..279S}   & 421 & 0.37 & 21:33:59.30 & +57:29:30.76\\

Cantat-Gaudin et al. (2018)   & 460 & 0.7 & 21:38:58.80 & +57:30:50.40 \\
\hline
\end{tabular}
\end{table*}

The identified member population towards this complex in the previous studies can broadly be divided into four categories. Spectroscopically identified members \citep{2002AJ....124.1585C,2006AJ....132.2135S,2013A&A...559A...3S}, {\it Spitzer} based NIR excess sources \citep{2004ApJS..154..385R,2006ApJ...638..897S,2009ApJ...702.1507M}, identification based on $\rm H_{\alpha}$ excess emission \citep{2011MNRAS.415..103B,2012AJ....143...61N}, and X-ray emission sources \citep{2007ApJ...654..316G,2009AJ....138....7M,2012MNRAS.426.2917G}. In addition, a relatively more recent analysis by \citet{2021AJ....162..279S} combines the near-infrared data from UKIRT with X-ray data from XMM-Newton to identify Class III YSO cluster members in a region covering the IC 1396A region. 
Altogether, there are 1791 candidate members identified in the literature. Apart from this, \citet{2018A&A...618A..93C} have analyzed a large number (1229) of Milky Way clusters using the Gaia-DR2 catalog. They used an unsupervised machine-learning technique to detect the member stars. They have listed the stars with membership probability greater than $50\%$ as candidate cluster members. For IC 1396, they have identified 460 stars within a radius of $~0.7^{\circ}$ centered at $\rm \alpha = 21:38:58.80$ and $\rm \delta = +57:30:50.40$. This region mostly covers the central part of the complex around the massive star HD 206267. Recently, \citet{2022arXiv221011930P}\footnote{This artcile is in press, hence detailed comparison of the sources could not be incorporated.}, using Gaia-EDR3 and optical spectroscopic analysis of the complex, provides distance, age, and distribution of the  the member sources. In Table \ref{work_mem}, we summarize details of the area covered and the number of stars retrieved in individual work.

We detect the member stars within the region of $1.5^\circ$ radius shown as a white dashed circle in Figure \ref{wise} and aim to detect new members of the complex. In Section \ref{compa_lit}, we compare the catalog identified in this work with the literature.

\section{Analysis \& results} \label{analy}
\subsection{Data from Gaia-DR3} \label{data}
To obtain the  Gaia-based membership of the region, we use the Gaia-DR3 catalog, downloaded from the Gaia archive\footnote{https://gea.esac.esa.int/archive/}. 
We retrieve all the sources within the $1.5^\circ$ radius centered at $\rm \alpha = 21:40:39.28$ and $\rm \delta = +57:49:15.51$. The search region is shown as the white dashed circle in Figure \ref{wise}, covering the entire IC 1396 complex. To identify the likely cluster members of this complex, we select sources based on the following criteria. All the selected sources must have positive parallax values ($\rm \pi>0~mas$). 
We consider all the sources with their proper-motion ranging between $\rm |\mu_{\alpha}cos\delta| \leq 20~mas/yr$ and $\rm |\mu_{\delta}| \leq 20~mas/yr$. This constraint on the proper motion values removes a large fraction of contaminants \citep{2018ApJ...869....9G, 2018AJ....156..121G}. All the sources we consider must-have magnitude values in G, BP, and RP bands. We thus obtain 458875 sources within the $1.5^\circ$ region which satisfy all the criteria mentioned above.

Following the histogram turnover method \citep{2007ApJ...669..493W,2013MNRAS.432.3445J,2017ApJ...836...98J,2017ApJS..229...28G,2021MNRAS.504.2557D}, we obtain the 90\% photometry completeness limits of G, BP, and RP bands to be 20.5, 21.5, and 19.5~mag, respectively. This is in agreement with the survey completeness, which is between $\rm G\approx 19$ and $\rm G\approx 21~mag$ \citep{2020arXiv201201533G}. The corresponding mass completeness limits are estimated in Section \ref{age}.

\subsection{Membership analysis} \label{memb_analysis}
Detecting the membership of a star-forming region is the first step towards analyzing its various star-formation properties. If the regions are large (e.g., IC 1396, Lupus) or the regions are not isolated, then the identification of members is not straightforward. Several authors have used different methods to achieve this. Here we briefly summarize the different methods of segregating the member stars from the field population. Pioneering works of \citet{1971A&A....14..226S, 1958AJ.....63..387V} adopt the probability measurements of stars using their proper motions to confirm their membership. In these works, they modeled the distribution of stars in the vector point diagram (VPD) using a bi-variate Gaussian mixture model (GMM). Later, adding the celestial coordinates of stars to their proper motions, \citet{1995AJ....109..672K} refined the membership probabilities. Some researchers selected the stars by partitioning data space into bins \citep{1991A&AS...87...69P,2012MNRAS.422.1495L}. In another work, \citet{2007A&A...470..585B} tried to separate the cluster members from the field stars based on their probability density in their VPD space. The broadband photometry is also considered as a tool to separate the cluster members from field stars with the help of color-magnitude (CMD) and color-color diagram (CCD) \citep{2004A&A...416..125D,2007A&A...470..585B}. 
\citet{2014A&A...561A..57K} have developed a method of computing membership probabilities in an unsupervised manner from the combination of celestial coordinates and photometric measurements. Their method is unsupervised photometric membership assignment in stellar clusters (UPMASK). The method of \citet{2014A&A...563A..45S,2019A&A...625A.115O} uses astrometric and photometric features of the stars for membership analysis. Then they apply the GMM with different components to model the field population and follow the Bayesian information criteria to choose a model. Then this method modeled the cluster with GMM in the astrometric space and a principal curve in the photometric space. Several recent works have used this methodology for membership analysis \citep{2020A&A...643A.148G,2021A&A...646A..46G}. The use of unsupervised and supervised computation of membership probabilities has also followed in several works in the recent past. In these works, the unsupervised GMM is used to generate a first catalog for the computation of supervised membership probability. These works used the random forest (RF; \citealt{breiman2001random, DBLP:journals/corr/abs-1201-0490}) classifier of the machine learning algorithm for the supervised computation of membership probability. Recently, \citet{2022arXiv220913302M} used various CMDs effectively along with RF classifier to obtain the membership of NGC 2244.

So both astrometry and photometric properties of stars play a crucial role in identifying member populations. As discussed, unsupervised and supervised membership probability estimation works efficiently and effectively. The crucial part of this method is preparing a training set, which comes through the unsupervised estimation, the GMM method. However, GMM suffers difficulty in filtration if the field contamination is relatively high. A safe way to overcome this difficulty is to combine the photometric properties in CMDs to obtain a set of stars, which can be used for the supervised membership probability estimation. Our present analysis uses the various CMDs to refine the member population obtained from the GMM method. We use a few CMDs and theoretical isochrones with prior knowledge about the nature of the star-forming complex from earlier studies. This helps to derive a cleaner member data set, which is used as a training set to derive the supervised membership probability using the RF classifier. We discuss the application of both GMM and RF in the following. More detail about the GMM method is explained in Appendix \ref{gmm_det}.

\subsubsection{Applying the Gaussian Mixture Model} \label{GMM}
We use five parameters (proper motions, parallax, and positions) for our clustering analysis using GMM. We have neither used the errors of the corresponding parameters nor the magnitude and color values as input parameters since they do not follow the Gaussian distributions. The GMM method fails drastically in cluster identification if we apply it to all stars (i.e., 458875 number of sources within the whole area). This is one of the significant limitations of the GMM method, which is also observed in other analyses \citep{2018AJ....156..121G,2018ApJ...869....9G}. The possible reasons for this failure are described by \citet{1990A&A...235...94C}. They pointed out that if the ratio between field stars and member stars is very high, it might cause an issue in clustering analysis using GMM. The other possible reason could be that the field stars do not follow a gaussian distribution.

To avoid the above issues related to the GMM method, we try to apply GMM over a small sample with minimum field star contamination. We must remember that obtaining the member population is not straightforward when dealing with a large star-forming complex such as IC 1396, whose radius is $\sim 1.5^\circ$. The reason is that the member populations of IC 1396 might not follow a single Gaussian distribution in their proper motion parameters, unlike an isolated cluster. So, we have to choose a small region very carefully, such that the astrometric and photometric properties of the stars in this region should represent the whole complex, and also, at the same time, the field star contamination should be as minimum as possible. In this work, we choose a conservative small central circular region of radius $30\arcmin$ around the coordinate mentioned in Section \ref{data}. We also use information from previous studies to minimize regional field star contamination. The previous studies suggest the distance of IC 1396 to be $\rm \sim 1~kpc$ \citep{2002AJ....124.1585C,2019A&A...622A.118S}. So we consider the stars that lie within the distance of 700~pc and 1100~pc to run the GMM algorithm so that we can safely throw the stars that lie outside the distance range. With these conditions, there are 6263 stars within the circular region of radius $30\arcmin$. We apply GMM on the 6263 stars, and based on the unsupervised membership estimation, we try to retrieve an initial sample of member stars, which will be used for the membership analysis based on supervised probability computation using the RF method.

\begin{figure}
\centering
\includegraphics[scale=0.45]{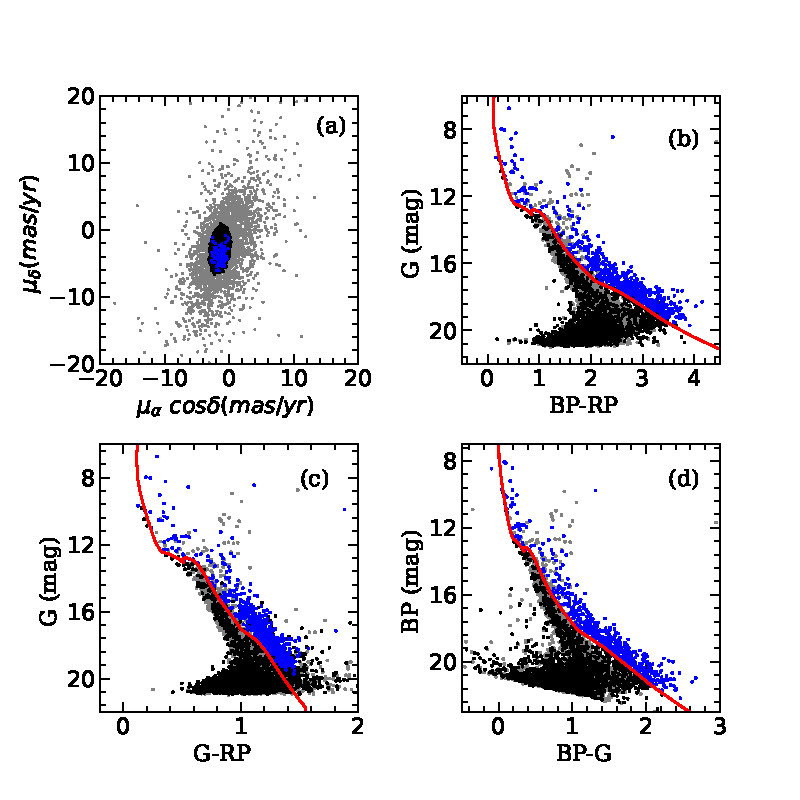}
\caption{(a) shows the proper-motion vector point diagram (VPD), and (b,c,d) show the various combination of CMDs of the 6263 sources within a $30\arcmin$ radius. On the CMDs, the red curve displays the PARSEC isochrone \citep{2014MNRAS.444.2525C} for 10~Myr, plotted after correcting for a distance of 900~pc and extinction of $\rm A_V=1~mag$ (discussed in section \ref{age}). In all the plots, the gray dots are 6263 stars, and the black dots are the 3760 stars separated by the GMM method with a probability greater than 80\%. Out of the 3760 stars, 577 stars lie right of 10~Myr  in all the CMDs are shown in blue dots. }
\label{gmm_vpd}
\end{figure}

Since the stars can broadly be separated into two groups as cluster members and field contaminants, we apply the GMM method with two components on these 6263 stars, and we retrieve 3760 stars with $\rm P_{GMM}\geqslant0.8$, and the remaining 2503 stars are mostly non-members consisting of the field star population. A few possible combinations of CMDs and the VPD of these 6263 stars (gray), along with the extracted 3760 starts (black) from GMM, are shown in Figure \ref{gmm_vpd}. As seen from the VPD diagram (Figure \ref{gmm_vpd}(a)), the 3760 stars populate as the central black region. This is expected since the member stars of a region usually lie within a narrow circular distribution in the VPD plot. However, the VPD plot and distribution of the 3760 stars on the CMDs show that the member stars are still associated with contamination. There could be a few probable reasons for this. In this analysis, we do not apply any constraint on the magnitude of stars to filter a maximum number of member stars in the fainter end. However, the fainter stars have higher uncertainty and are less reliable. The other possible reason is that in the case of a giant star-forming region such as IC 1396, the member stars might have a little wide distribution in proper motions compared to an isolated stellar cluster. That again increases the chance of contamination in the member star population. So it requires a double check to minimize contamination from the 3760 stars extracted from the GMM method. For this, we use various CMDs, shown in Figure \ref{gmm_vpd}.
Though the cluster associated with IC 1396 have a mean age of $\sim$ 2$-$4~Myr \citep{2005AJ....130..188S}, but there is a spread in age up to 10~Myr for some stars, so here we consider only those sources younger than 10~Myr as members. This further removes a significant fraction of contaminated stars from the member population. There are 577 stars left which are more reliable to be member stars. These 577 stars are shown as blue dots in Figure \ref{gmm_vpd}. 
The selected member stars show a distribution that is largely indistinguishable from the field stars,  likely due to the large number of field stars along the line of sight compared to the small number of cluster stars. However, compared to the distribution of filed stars, the distribution of member stars peaks at different locations and shares conservative space in the VPD diagram. For a training sample for the RF method, we keep the 577 stars as member stars and the 2503 stars as non-member stars.

\subsubsection{Applying the random forest classifier method} \label{RF}
In this section, we apply the supervised machine-learning technique, RF classifier to identify the membership of the entire complex.
This technique is an ensemble of machine-learning decision trees for classification and regression tasks. Due to its robustness, the RF technique is widely used in the astrophysical field \citep{2011MNRAS.414.2602D,
2013MNRAS.435.1047B,
2017ApJ...843..104L,2018PASJ...70S..39L,2018MNRAS.476.3974P,
2018AJ....156..121G,2018ApJ...869....9G,2021arXiv210305826M}. In this work, we use the python-based RF classifier available in the scikit-learn package\footnote{https://scikit-learn.org/stable/modules/generated/sklearn.ensemble.\\
RandomForestClassifier.html}.

Before using RF on the total population to identify member stars, we need to train the machine, as described in Appendix \ref{rf_eff}. After checking RF's efficiency, we run the RF method to obtain the most probable population of the whole complex. The relative importance of the parameters in separating the member and non-member stars is also listed in Appendix \ref{rf_eff}. After training the machine with the training set retrieved from the GMM method, we ran the RF classifier on a total of 458875 stars located in the direction of the complex IC 1396. Out of these stars, we need to retrieve the most reliable member population of the complex. As described in Appendix \ref{rf_eff}, while training the machine, a few color and magnitude terms also become essential in segregating members from non-member populations. In order to make the detection more robust, we can use the parallax parameter to filter out the non-member stars. Here, we run RF on the stars ($\sim 70000$), which lie within the parallax range of (0.8 to 1.6~mas). With this, we can use the color and magnitude parameters effectively; otherwise, this could increase more unlikely sources.

\begin{figure}
\centering
\includegraphics[scale=0.42]{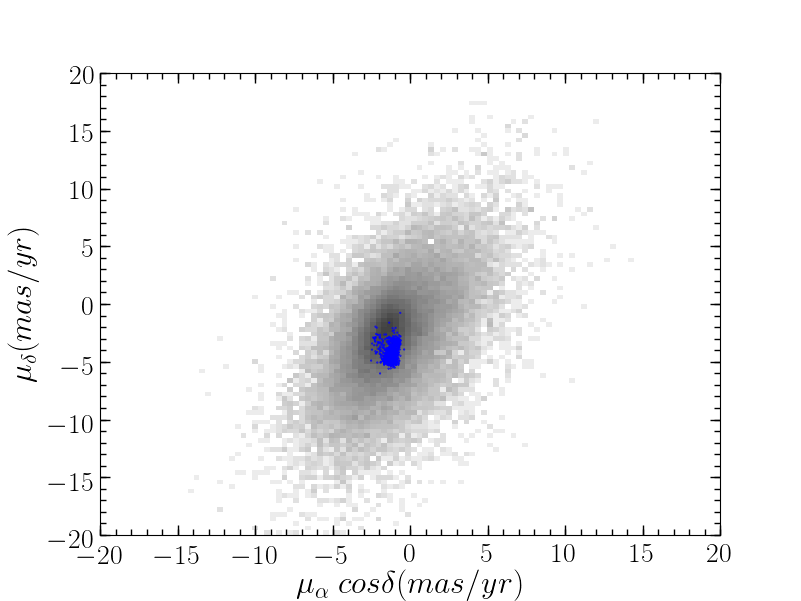}
\caption{VPD of all the 458875 stars is shown as a gray density plot. The blue dots indicate the member population with $\rm P_{RF}\geqslant0.8$.}
\label{rf_vpd}
\end{figure}

RF provides a membership probability to each star based on its training in the previous step. In our analysis, we retrieve 1803 likely possible members with a probability value of $\rm P_{RF}\geqslant0.6$. Details of the 1803 likely member stars are listed in Table \ref{star_list}. Of these 1803 stars, 1243 have a high probability value of $\rm P_{RF}\geqslant0.8$. Hereafter we use these highly probable candidate members for follow-up analysis. In this work, the massive star HD 206267 has $\rm P_{RF}<0.6$. 
HD 206267 is an multiple-star system of spectral type O5V $-$ O9V and an older member of the cluster \citep{2012A&A...538A..74P,2020A&A...636A..28M}. RUWE, parallax, $\rm \mu_{\alpha}cos\delta$, and $\rm \mu_{\delta}$ of the star are 5.07, 1.360$\pm$0.218~mas, -1.951$\pm$0.120, and -5.493$\pm$0.281~mas/yr, respectively. The multiple stellar systems resulted in higher RUWE values and proper motions. The radial velocity of the star is $\rm -24.8\pm1.4~ km/s$ \citep{2021ApJS..254...42B}, which is well within the radial velocity distribution of the member stars (Figure \ref{hist_Rv}). This is a direct confirmation of its membership. Also, many earlier studies using multi-wavelength data sets show the connection of the massive star with the star-forming complex \citep{1995ApJ...447..721P,2012MNRAS.426.2917G,2014A&A...562A.131S,2015A&A...573A..19S,2019A&A...622A.118S}.
In Figure \ref{rf_vpd}, we plot the proper motion VPD plot for all the 458875 stars. The member stars with $\rm P_{RF}\geqslant0.8$ identified by RF are shown as blue dots. This plot shows that the members are concentrated within a narrow range of proper motion values.

\begin{figure}
\centering
\includegraphics[scale=0.35]{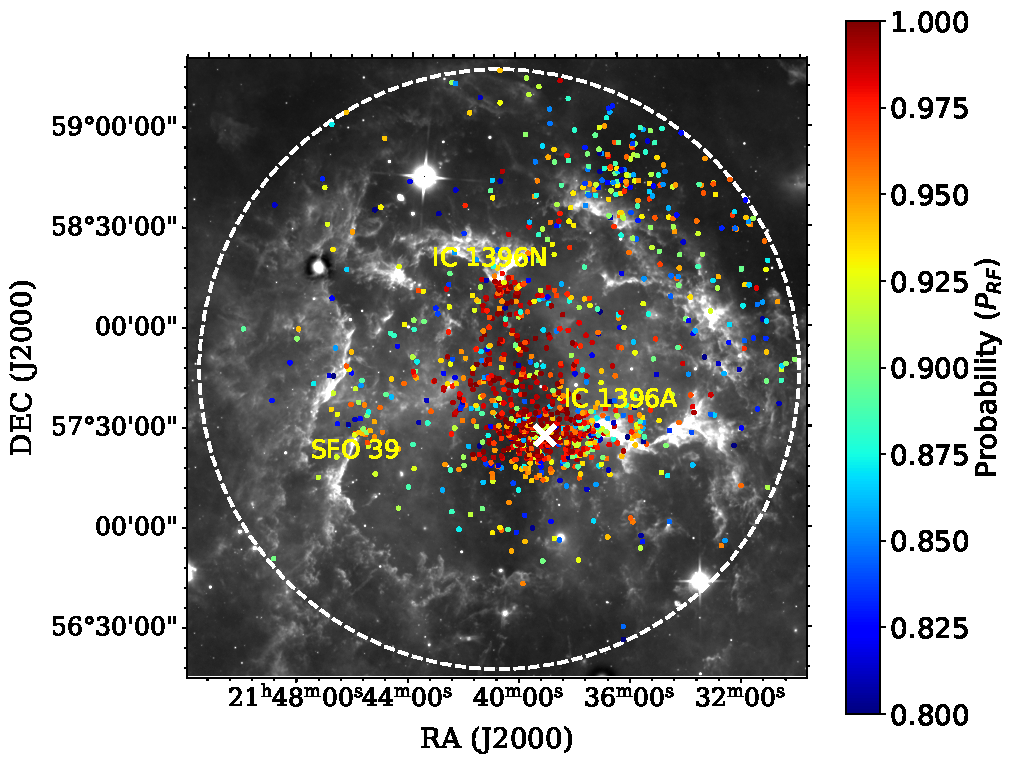}
\caption{Spatial distribution of the likely cluster members identified from the RF method on the WISE $\rm 22~\mu m$ band. The white `$\times$' symbol marks the position of the massive central star HD 206267. The candidate members' color code is based on their $\rm P_{RF}$ values, and their color bar is also shown. Locations of the BRCs IC 1396A, IC 1396N, and SFO39 are labeled on the plot.
 }
\label{rf_wise}
\end{figure}

The G versus G-RP CM diagram is shown in Figure \ref{g_rp_cm_iso} for the member stars ($\rm P_{RF}\geqslant0.8$) within the region of radius $1.5^\circ$ (shown in Figure \ref{wise}). All the identified member stars indicate a well-defined pre-main-sequence locus on the CM diagram.  
In Figure \ref{rf_wise}, we over-plot the likely members on the $\rm 22~\mu m$ WISE image, highlighting their distribution as a function of their $\rm P_{RF}$ value. An over-density of the source distribution is visible in the central part of IC 1396. Within the complex, the stars display a diagonal distribution ranging from the BRC IC 1396A to the IC 1396N. Most of the stars are clustered around the massive star HD 206267, shown as the white `$\times$' symbol in the figure. IC 1396N is also associated with a small cluster. A tiny clustering of stars is also visible towards the tip of BRC SFO39. A small fraction of stars is also seen to be randomly distributed all around the complex. A clustering of stars also found towards the northern periphery of the complex. The overall distribution of stars is higher towards the west than the east of the complex.

\begin{table*}
\small
\centering
\caption{List of the Gaia-based member population identified using the RF method. The table provides the positions, parallax, proper motions, and magnitude values in G, BP, and RP bands along with $\rm P_{RF}$ values of 1803 stars identified with $\rm P_{RF}\geqslant0.6$. For analysis in this paper, we consider stars with $\rm P_{RF}\geqslant0.8$. }
\label{star_list}
\begin{tabular}{ccccccccccc}
\\ \hline

Star No. & RA (2000) & DEC (2000) & RUWE & Parallax & pmra & pmdec & G & BP & RP & $\rm P_{RF}$\\
         & (degree)  & (degree)   &      & (mas)    & (mas/yr) & (mas/yr) & (mag) & (mag) & (mag) & \\
\hline
1 & 327.6486 & 57.3557 &  0.922 & 1.204$\pm$0.178 & -2.553$\pm$0.232 & -3.151$\pm$0.194 & 18.80 & 20.60 & 17.56  & 0.716 \\ 
2 & 327.6364 & 57.4715 &  1.055 & 1.299$\pm$0.157 & -2.162$\pm$0.194 & -3.075$\pm$0.156 & 18.59 & 20.34 & 17.33  & 0.788 \\ 
3 & 327.6372 & 58.4742 &  0.951 & 0.940$\pm$0.012 & -1.580$\pm$0.016 & -3.890$\pm$0.012 & 12.04 & 13.02 & 11.07  & 0.682 \\ 
4 & 327.4235 & 57.5640 &  0.999 & 1.032$\pm$0.022 & -3.746$\pm$0.027 & -4.302$\pm$0.024 & 15.18 & 15.95 & 14.30  & 0.682 \\ 
5 & 327.3769 & 57.5977 &  1.043 & 1.009$\pm$0.013 & -4.326$\pm$0.016 & -4.777$\pm$0.014 & 11.15 & 11.53 & 10.57  & 0.664 \\ 
6 & 327.4919 & 57.6289 &  0.964 & 1.018$\pm$0.011 & -3.333$\pm$0.012 & -2.184$\pm$0.011 & 12.53 & 13.71 & 11.46  & 0.682 \\ 
7 & 327.1520 & 57.5298 &  0.962 & 1.017$\pm$0.014 & -3.234$\pm$0.017 & -4.564$\pm$0.015 & 14.04 & 14.82 & 13.16  & 0.780 \\ 
8 & 327.2893 & 57.6506 &  0.852 & 1.094$\pm$0.012 & -2.117$\pm$0.014 & -4.260$\pm$0.012 & 11.20 & 11.44 & 10.80  & 0.828 \\ 
9 & 327.5741 & 57.8185 &  0.981 & 1.025$\pm$0.069 & -1.073$\pm$0.076 & -3.113$\pm$0.082 & 17.59 & 19.09 & 16.42  & 0.702 \\ 
10 & 327.1886 & 57.7081 &  1.014 & 1.049$\pm$0.030 & -1.048$\pm$0.035 & -4.333$\pm$0.033 & 15.98 & 17.29 & 14.84  & 0.766 \\ 
\hline
\end{tabular}
\\(This table is available in its entirety as online material. A portion is shown here for guidance regarding its form and content.)
\end{table*}

\subsection{Characterstics of the member stars} \label{charc_stars}
\begin{figure}
\centering
\includegraphics[scale=0.5]{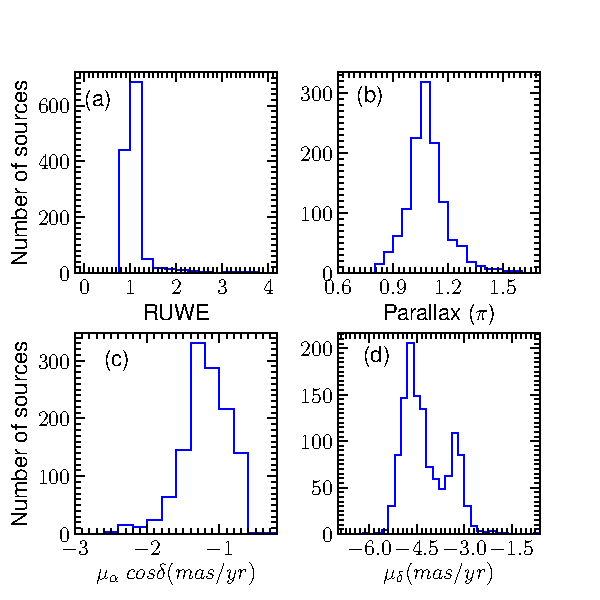}
\caption{Histogram distributions of RUWE (a), parallax (b),  $\rm \mu_{\alpha}cos\delta$ (c), and  $\rm \mu_{\delta}$ (d) of the 1243 member stars identified in this work. Bin size of the histograms are 0.25, 0.05~mas, 0.2~mas/yr, and 0.2~mas/yr, for RUWE, parallax, $\rm \mu_{\alpha}cos\delta$, and $\rm \mu_{\delta}$, respectively. }
\label{hist_parallax}
\end{figure}

In Figure \ref{hist_parallax}, we show histogram distributions of RUWE\footnote{$\rm https://gea.esac.esa.int/archive/documentation/GDR2/ \\
Gaia\_archive/chap\_datamodel/sec\_dm\_main\_tables/ssec\_dm\_ruwe.html$} (Renormalised unit weight error), parallax, and proper-motions of member stars detected in this work. Table \ref{gaia_range} provides the range of these parameters. RUWE parameter provides a measure of astrometric solutions. The RUWE value of around 1.0 is expected for sources where the single-star model provides a good fit for the astrometric observations. Stars with RUWE greater than 1.4 are considered resolved doubles \citep{2020arXiv201201533G}. In our list of selected members, only 144 and 82 stars have RUWE $>$ 1.4, from the list with $\rm P_{RF}\geqslant0.6$, and 0.8, respectively. 
These sources with higher RUWE could be multiple-star systems. The stars detected in this work are of good quality sources. Out of the 1243 stars, $\sim95\%$ stars have relative parallax error less than $20\%$.

\begin{table}
\tiny
\centering
\caption{Range, mean, median, and standard deviation of RUWE, parallax, and proper motions of the 1243 member stars.}
\label{gaia_range}
\begin{tabular}{ccccc}
\\ \hline \hline
Parameter & Range  & Mean & Median & SD \\
\hline
RUWE                                 & $0.77 - 13.79$  & 1.12 & 1.02 & 0.59  \\
Parallax (mas)                       & 0.834$\pm$0.162 -- 1.564$\pm$0.184 & 1.085$\pm$0.003 & 1.078 & 0.109    \\	
$\rm \mu_{\alpha}cos\delta$ (mas/yr) & -2.506$\pm$0.006 -- -0.378$\pm$0.015 & -1.194$\pm$0.002 & -1.187 & 0.325  \\
$\rm \mu_{\delta}$ (mas/yr)          & -6.011$\pm$0.216 -- -0.764$\pm$0.014 & -4.215$\pm$0.004 & -4.404 & 0.712  \\
\hline
\end{tabular}
\end{table}

Figure \ref{hist_parallax} (b) displays the histogram distribution of parallaxes for all these identified member stars. In parallax, the stars detected in this work lie within a spread of $\sim$0.8~mas with mean, median, and standard deviation values of $\rm 1.085\pm0.003~mas$, 1.078~mas, and 0.109~mas, respectively. The distance to the cluster is estimated using the parallax values of those sources whose relative parallax error ($\sigma \pi/ \pi$) is better than 20\% and $\rm RUWE<1.4$. 
Out of 1243, we find 1107 stars satisfy this condition. From these 1107 stars, we estimate the weighted mean parallax to be $\rm 1.090\pm0.003~mas$, which translates to a distance of $\rm 917\pm2.7~pc$. This distance estimate matches well with earlier estimates in literature \citep{2002AJ....124.1585C, 2019A&A...622A.118S,2022arXiv221011930P}.

In Figure \ref{hist_parallax} (c) and (d), we show the histogram distributions of the proper motions ($\rm \mu_{\alpha}cos\delta, and ~\mu_{\delta}$). We derive the mean, median, and standard deviation values for $\rm \mu_{\alpha}cos\delta$ to be $\rm -1.194\pm0.002~mas/yr$, $\rm -1.187~mas/yr$, and $\rm 0.325~mas/yr$, respectively. For $\rm \mu_{\delta}$, these values are $\rm -4.215\pm0.004~mas/yr$, $\rm -4.404~mas/yr$, and $\rm 0.712~mas/yr$, respectively.

\subsection{Comparision with literature} \label{compa_lit}
In this section, we compare our detected member stars with the sources detected in the literature. As discussed in Section \ref{lit_stars}, there are 1791 stars detected towards the complex based on various surveys. Also, using Gaia-DR2 data, \citet{2018A&A...618A..93C}, detected 460 stars towards IC 1396. We compare our findings separately with the source lists found in the literature.

To compare with the sources of various surveys, we first find their Gaia-DR3 counterpart information. Out of the 1791 stars, 1002 stars have Gaia counterparts. Then we refine the catalog further based on the astrometry quality. Thus we use the 705 stars, which have a relative parallax error of $<20\%$, for comparison. Of the 705 stars, 360 stars ($\sim51\%$) are retrieved in our work as member stars with $\rm P_{RF}\geqslant0.8$. The number is 409 ($\sim 60\%$) with $\rm P_{RF}\geqslant0.6$. Due to their poor membership probability, the remaining stars are not detected as members. 

Then we compare our member list with the 460-star list of \citet{2018A&A...618A..93C}. Within the common area, out of the 460 stars, we retrieved 348 ($\sim76\%$) stars in this work with $\rm P_{RF}\geqslant0.8$. The number is 389 ($\sim85\%$) with $\rm P_{RF}\geqslant0.6$.
In this work, we consider only the stars with a higher probability of $80\%$. In \citet{2018A&A...618A..93C}, they considered all the stars with membership probability above $50\%$. So the stars, with higher probability, are retrieved in our work. In our work, we identify more member stars than \citet{2018A&A...618A..93C} mainly due to the large area we consider.

\begin{figure}
\centering
\includegraphics[scale=0.5]{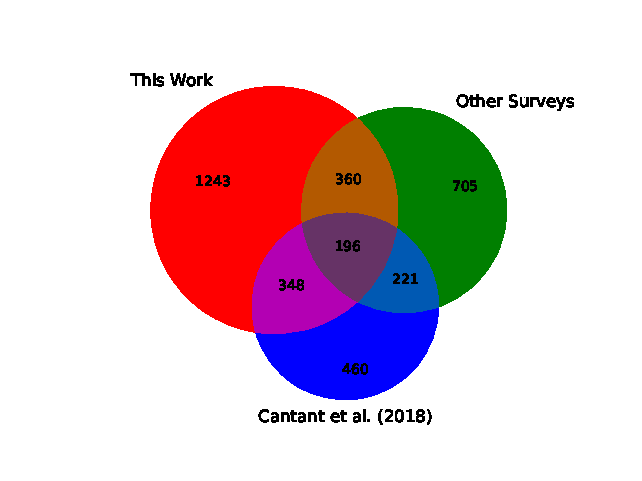}
\caption{Venn diagram summarizing the comparison between the member population from this work with the stars from several other surveys, and with stars from work of \citet{2018A&A...618A..93C}. }
\label{venn_compa}
\end{figure}

Then we also compared the source list obtained by the various surveys (section \ref{lit_stars}) with the stars detected by \citet{2018A&A...618A..93C}. Here, also we considered the good quality 705 stars for comparison. In this case, we found 221 ($\sim31\%$) survey-based stars common with the catalog of \citet{2018A&A...618A..93C}. There are 196 stars common to all three catalogs discussed here. We summarize the analysis as a Venn diagram (Figure \ref{venn_compa}).

\section{Properties of the complex} \label{res}
\subsection{Sub-clusters within the complex} \label{cluster}
\begin{figure}
\centering
\includegraphics[scale=0.35]{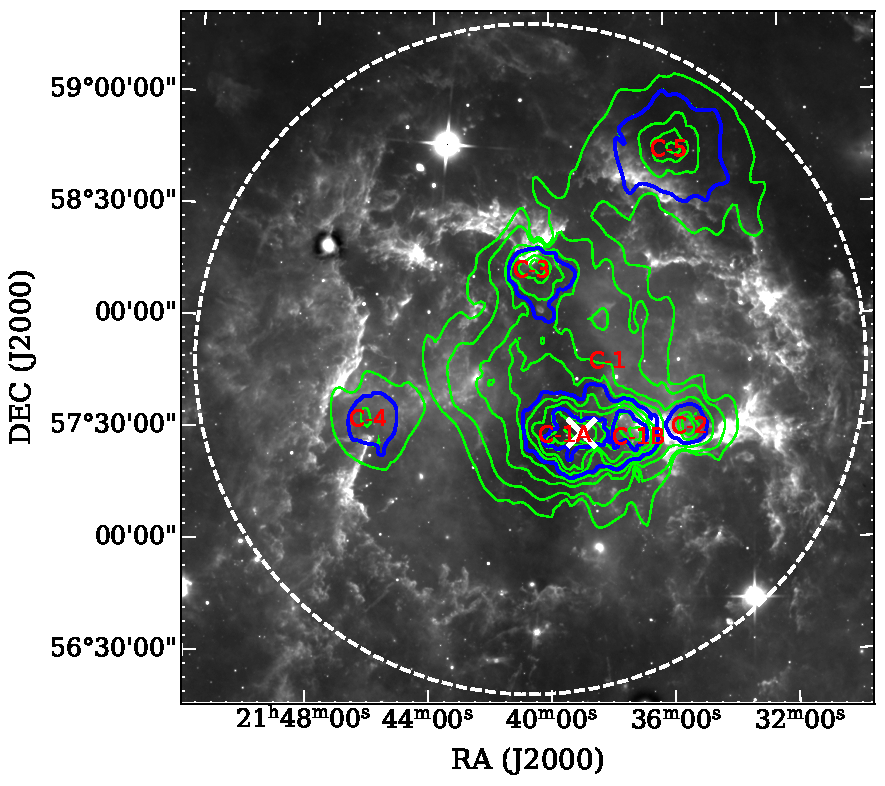}
\caption{Contours of stellar surface density distribution generated using the 1243 candidate members identified towards IC 1396 complex overlaid on the WISE $\rm 22~\mu m$ band image. Contours are at levels of 0.6, 1, 2, 3, 5, 7, 10, 12, and 20 $\rm stars~ pc^{-2}$. The white `$\times$' symbol marks the position of the massive star HD 206267. Different clusters are retrieved from the stellar density map. The blue curves are the clusters shown along with their nomenclature.}
\label{NN}
\end{figure}

The spatial distribution of the 1243 stars (Figure \ref{rf_wise}) displays the association of clustering with IC 1396. In this section, we attempt to identify the clusters quantitatively. To do this, we generate the surface density plot using the 1243 member stars and apply the nearest neighbor (NN) method \citep{1985ApJ...298...80C,2011AN....332..172S}. According to this method, the j-th nearest neighbour density is defined as

\begin{equation}
\rm \rho_j~ = ~ \frac{j-1}{S(r_j)}
\end{equation}

where $\rm r_j$ is the distance to its j-th nearest neighbour and $\rm S(r_j), $ is the surface area with radius $\rm r_j$.  To obtain the distribution of member stars, we use $\rm j = 20$, which is found to be an optimum value for cluster identification \citep{2008MNRAS.389.1209S,2017MNRAS.465.4753R,2021MNRAS.504.2557D}. With this procedure, we generate the stellar density map with a pixel size of 0.1~pc ($20.5^{\prime\prime}$). Figure \ref{NN} shows the WISE 22~$\rm \mu m$ map overlaid with density contours. The lowest contour is at 0.6~stars $\rm pc^{-2}$, within which the maximum number of sources falls. These stellar density contours reveal the cluster of stars towards the star-forming complex.

\begin{figure*}
\centering
\includegraphics[scale=0.5]{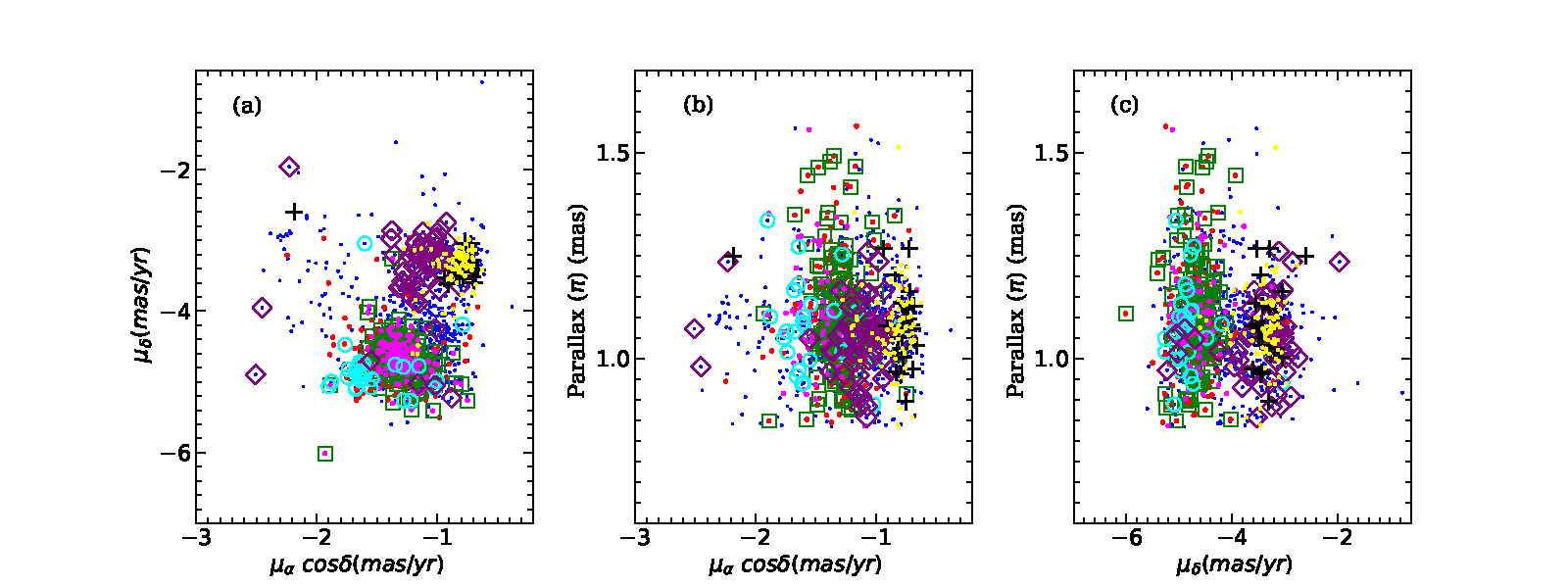}
\caption{The spatial distribution of proper motions and parallax of the member stars detected in this work. Blue dots represent the 1243 member stars. The population of the clusters is shown in different colors and shapes. Cluster C-1 (red dot), C-1A (green square), C-1B (magenta dot), C-2 (cyan circle), C-3 (purple diamond), C-4 (black plus mark), and C-5 (yellow dot).}
\label{parallax_propermotion}
\end{figure*}

\begin{table*}
\tiny
\centering
\caption{The number of stars, mean, median, and standard deviation of RUWE, parallax, and proper motions of the clusters associated with IC 1396.}
\label{stat_cluster}
\begin{tabular}{ccccccccccccccc}
\\ \hline \hline
Cluster & Radius & No. & \multicolumn{3}{c}{RUWE} & \multicolumn{3}{c}{Parallax} & \multicolumn{3}{c}{$\rm \mu_{\alpha}cos\delta$} & \multicolumn{3}{c}{$\rm \mu_{\delta}$} \\
     & (pc)  & of stars   & \multicolumn{3}{c}{} & \multicolumn{3}{c}{(mas)} & \multicolumn{3}{c}{(mas/yr)} & \multicolumn{3}{c}{(mas/yr)} \\
\hline
     &    &  & Mean & Median & SD & Mean & Median & SD & Mean & Median & SD & Mean & Median & SD \\
\hline
C-1  & 3.80 & 426 & 1.11 & 1.02 & 0.38 & 1.098$\pm$0.006 & 1.084 & 0.118 & -1.329$\pm$0.004 & -1.327 & 0.197 & -4.671$\pm$0.006 & -4.691 & 0.334 \\
C-1A & 1.72 & 162 & 1.11 & 1.02 & 0.37 & 1.101$\pm$0.009 & 1.079 & 0.126 & -1.298$\pm$0.007 & -1.309 & 0.159 & -4.690$\pm$0.010 & -4.699 & 0.287 \\
C-1B & 1.22 & 80 & 1.12 & 1.02 & 0.53 & 1.108$\pm0.013$ & 1.103 & 0.116 & -1.442$\pm$0.010 & -1.449 & 0.222 & -4.656$\pm$0.016 & -4.656 & 0.366 \\
C-2 & 1.40 & 27 & 1.07 & 1.03 & 0.21 & 1.076$\pm$0.026 & 1.066 & 0.102 & -1.512$\pm$0.019 & -1.608 & 0.274 & -4.825$\pm$0.030 & -4.882 & 0.411 \\
C-3 & 2.17 & 60 & 1.11 & 1.02 & 0.38 & 1.047$\pm$0.013 & 1.052 & 0.085 & -1.172$\pm$0.008 & -1.117 & 0.317 & -3.438$\pm$0.014 & -3.266 & 0.580 \\
C-4 & 1.84 & 23 & 1.04 & 1.02 & 0.09 & 1.085$\pm$0.021 & 1.080 & 0.097 & -0.831$\pm$0.014 & -0.772 & 0.296 & -3.332$\pm$0.028 & -3.417 & 0.225 \\
C-5 & 3.76 & 87 & 1.07 & 1.02 & 0.12 & 1.083$\pm$0.012 & 1.074 & 0.095 & -0.854$\pm$0.007 & -0.816 & 0.141 & -3.401$\pm$0.013 & -3.330 & 0.336 \\
\hline
\end{tabular}
\end{table*}

For identification of the clusters in this region, we use the {\it astrodendro} algorithm \citep{2019ascl.soft07016R} in Python. This algorithm works based on constructing tree structures starting from the brightest pixels in the dataset and progressively adding fainter and fainter pixels. It requires the threshold flux value (minimum value), contour separation (min delta), and the minimum number of pixels required for a structure to be considered a cluster. In our analysis, we use the threshold and minimum delta to be 1.0, 0.3~stars $\rm pc^{-2}$, respectively. We use the minimum number of pixels as 150 to detect the potential clusters. These parameters are adopted after multiple trials for optimal detection of clusters. We identify six individual leaf structures with these input parameters, which we call clusters here. Two individual clusterings (C-1A and C-1B) are seen towards the massive star HD 206267, and collectively (C-1) is the central cluster of this complex. Towards the tail of BRC IC 1396A, another grouping (C-2) of stars is also seen. Except for this cluster towards the central part, another three clusters are also seen. They (C-3 and C-4) are linked with the BRC IC 1396N and SFO 39, respectively. We also detect a cluster (C-5) close to the boundary of the star-forming complex. Cluster identification in our work matches well with the clusters identified by \citet{2012AJ....143...61N} from the $\rm H_{\alpha}$ emission line survey. In their work, a cluster is associated with the southern BRC SFO 37. However, in our analysis, we cannot see any such cluster with SFO 37, which could be due to the sensitivity of Gaia. The cluster (C-5), which we detect in this work, was not seen by \citet{2012AJ....143...61N}, which could be because, their work survey a lesser area than the area covered in this study.

In Table \ref{stat_cluster}, we list the statistics (radius, number of stars, mean, median, and standard deviation) of RUWE, parallax, $\rm \mu_{\alpha}cos\delta$, and $\rm \mu_{\delta}$ for all the identified clusters. We derive physical radius ($\rm R_{cluster} = (A_{cluster}/\pi)^{0.5}$; \citealt{2017MNRAS.472.4750D}) of the clusters using the apertures retrived from {\it astrodendro}. Area of each cluster is calculated as $\rm A_{cluster}=N\times A_{pixel}$, where N is the number of pixels and $\rm A_{pixel}$ is the area of each pixel. The distribution of parallax and the proper motions of the cluster stars are also displayed in Figure \ref{parallax_propermotion}. We see from this plot two groupings. 
As we see, one is the larger group, mainly from the stars of the C-1 and C-2 clusters, and the second is a smaller group that appears due to stars from the other three (C-3, C-4, and C-5) clusters. This is also evident from the histogram distribution of the $\rm \mu_{\delta}$ (Figure \ref{hist_parallax}(d)). To quantitatively confirm our findings, we carry out a two-component Kolmogorov-Smirnov (KS) test with parallax and proper motions. The $p-$ score from the test is minimal and close to zero for the proper motions. For parallax, the $p-$ score is 0.02. This quantitatively confirms that proper motion parameters are the distinctive astrometric features, distinguishing the stars projected in the two sub-groups, which is seen in Figure \ref{parallax_propermotion}.

\subsection{Age and mass range of the candidate cluster members}\label{age}
\begin{figure}
\centering
\includegraphics[scale=0.5]{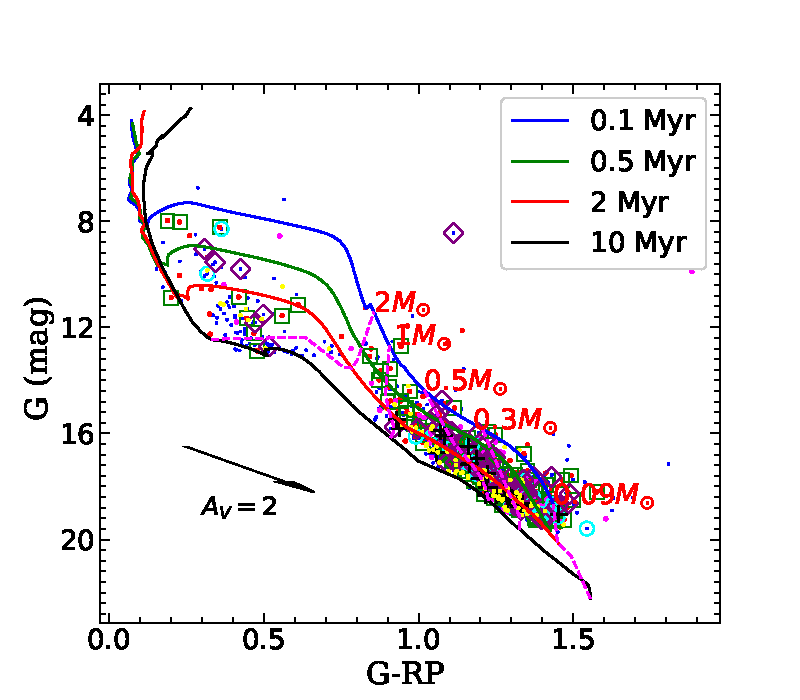}
\caption{G versus G-RP CMD of the 1243 member stars within the IC 1396 complex. PARSEC isochrones of 0.1, 0.5, 2.0, and 10.0~Myr are over-plotted. All the curves are plotted after correcting the distance (917~pc) and minimum extinction ($\rm A_V=1~mag$) (see text for details). Evolutionary tracks for stars having the mass of 0.09, 0.3, 0.5, 1.0, and 2.0~$\rm M_{\odot}$ are also shown. The colored symbols have the same meaning as in Figure \ref{parallax_propermotion}.}
\label{g_rp_cm_iso}
\end{figure}

\begin{figure}
\centering
\includegraphics[scale=0.4]{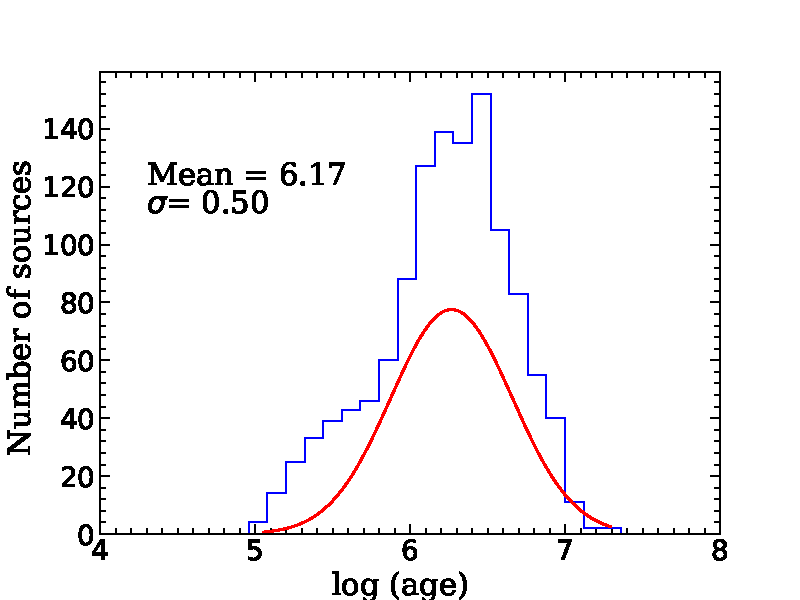}
\caption{Histogram distribution of the logarithmic age of the 1243 candidate members of IC 1396. The red curve displays the Gaussian fit.}
\label{hist_age}
\end{figure}

In this section, we estimate the mean age and mass completeness limit of the member stars identified in this analysis.  
Studies like \citet{2005AJ....130..188S,2012MNRAS.426.2917G} and references therein claim an approximate age of $\sim$4~Myr for the primary cluster.
To estimate the member population's age and mass completeness limit, we use the PARSEC isochrones available for the filters of Gaia-DR3 \citep{2014MNRAS.444.2525C}. We need to correct the isochrones for distance and extinction to fit them. 
In an earlier study using NIR and optical data, \citet{2005AJ....130..188S} have derived the average visual extinction value towards the entire complex to be $\rm A_V = 1.5\pm0.5~mag$. This value also matches the estimations by \citet{2002AJ....124.1585C} and \citet{2012AJ....143...61N}. 
The majority of detected stars in this work are located towards the central part of IC 1396, which is expected to be of less extinction due to the presence of massive star(s) around them compared to the surrounding regions such as BRCs, which are associated with the dense molecular clouds. For further analysis, we use the minimum extinction value of $\rm A_V = 1~mag$ obtained from \citet{2012AJ....143...61N}.

After correcting for distance (917~pc) and extinction ($\rm A_V = 1~mag$), we plot the isochrones of various ages on the G versus G-RP CMD in Figure \ref{g_rp_cm_iso}. To correct the extinction in individual bands for all the sources, we use the empirical relations of $\rm A_G/A_V,  A_{RP}/A_V$ \citep{2018A&A...616A..10G,2019A&A...623A.108B}. In Figure \ref{g_rp_cm_iso}, we plot various isochrones of evolutionary ages 0.1, 0.5, 2, and 10~Myr along with the evolutionary tracks corresponding to 0.09, 0.3, 0.5, 1, and 2~$\rm M_{\odot}$. From Figure \ref{g_rp_cm_iso}, we derive the age of individual stars by assigning the age of the closest isochrone. Similarly, by assigning the closest mass evolutionary track, we derive the mass of individual stars. However, local variation in extinction and binarity of stars might affect the accurate estimation of these parameters. In Figure \ref{hist_age}, we show the histogram distribution of logarithmic values of the age. By fitting a Gaussian curve to the distribution, we obtain the mean logarithmic age of the cluster to be $6.17\pm0.50$, which corresponds to a mean age of $\sim1.5\pm1.6$~Myr. 
Using of the upper limit of extinction, i.e., $\rm A_V = 1.5~mag$, the mean age obtained to be $\sim1.6\pm1.7$~Myr, which is still in match with the previous studies.

As discussed in Section \ref{data}, we see that the 90\% completeness limits of G, BP, and RP bands are 20.5, 21.5, and 19.5~mag, respectively. We use the G-band to estimate the mass-completeness limit of the cluster. Using an extinction value of $\rm A_V = 1 - 1.5~mag$, distance of 917~pc and considering  pre-main-sequence isochrone of 2~Myr \citep{2014MNRAS.444.2525C}, the magnitude limit of G-band (20.5~mag) corresponds to a mass of $\rm \sim 0.1 - 0.2 ~M_{\odot}$. This analysis shows that the Gaia-DR3 is complete down to the low-mass end. However, compared to the central region, i.e., towards the IC 1396A region, the extinction might be higher due to the presence of BRC and an associated molecular cloud. 
This local variation in extinction will play a role in the local mass completeness of the member stars towards the outer edge of the complex.

We list the mean, median, and standard deviation values of log(age) and mass for the entire complex and the individual clusters in Table \ref{stat_age_mass}. The mean and median values of log(age) and mass are similar, considering the whole complex and the clusters. This suggests that most of the population has evolved within the similar time scale of $\rm \sim~ 3~Myr$. However, previous studies have shown that, in the proximity of BRC candidates, multi-episodic star-formation is happening \citep{2014A&A...562A.131S}. Similarly, the stellar mass distribution appears uniform for the entire complex, which can be seen from the mean and median values for all the clusters. However, local mass segregation might be happening within the individual clusters.

\begin{table}
\small
\centering
\caption{Mean, median, and standard deviation for log(age) and mass derived from stars of whole IC 1396 and for the clusters.}
\label{stat_age_mass}
\begin{tabular}{ccccccc}
\\ \hline \hline
Cluster & \multicolumn{3}{c}{log(age)} & \multicolumn{3}{c}{Mass}\\
     & \multicolumn{3}{c}{(yr)} & \multicolumn{3}{c}{($\rm M_{\odot}$)}\\
\hline
     &  Mean & Median & SD & Mean & Median & SD  \\
\hline
Full & 6.17 & 6.25 & 0.49 & 0.68 & 0.43 & 1.01 \\
C-1  & 6.05 & 6.14 & 0.48 & 0.62 & 0.4 & 1.07 \\
C-1A & 6.06 & 6.14 & 0.49 & 0.70 & 0.4 & 1.37 \\
C-1B & 5.98 & 6.07 & 0.47 & 0.60 & 0.37 & 0.93 \\
C-2  & 5.99 & 6.20 & 0.40 & 0.66 & 0.35 & 1.29 \\
C-3  & 5.97 & 5.98 & 0.48 & 0.74 & 0.36 & 1.26 \\
C-4  & 6.33 & 6.34 & 0.30 & 0.44 & 0.45 & 0.20 \\
C-5  & 6.37 & 6.42 & 0.37 & 0.70 & 0.50 & 0.70 \\
\hline
\end{tabular}
\end{table}

\subsection{Cluster properties} \label{clust_prop}
Several clusters have been identified towards IC 1396 based on the spatial distribution of the associated stellar members. Each cluster leaves an imprint of the ongoing star formation in the complex. In this section, we briefly discuss the formation of clusters taking into account their age and spatial distribution.

\subsubsection{Inner clusters (C-1 and C-2)} \label{c1_c2}
Clusters (C-1 and C-2) are located towards the center of the complex. 
Also, two sub-clustering (C-1A and C-1B) are observed within cluster C-1.  
Subcluster C-1A is on the eastern side, and C-1B is on the western side of the massive star. The subcluster C-1B is linked to the head of BRC IC 1396 A, while the C-2 is seen towards its tail. C-1A contains more stars with slightly higher ages than C-1B. So the mean age of C-1A is slightly higher compared to C-1B.
Similarly, the mean age of cluster C-2 is similar to C-1B. This indicates a multi-generation star formation triggeded by the feedback effect of the central massive star. Earlier studies \citep{2014A&A...562A.131S,2019A&A...622A.118S,2022arXiv221011930P} have reported such triggered star formation activities towards the head of IC 1396 A. The presence of cluster C-2 is also a signature of ongoing triggered star formation towards the BRC complex. Using {\it Herschel} PACS images and analyzing the properties of young members in the head of IC 1396 A, \citep{2014A&A...562A.131S} suggested that this second generation of star-formation is triggered via radiative driven implosion (RDI) induced by the massive star HD 206267. However, more in-depth analysis with multiwavelength data would be helpful to understand the mechanism behind the triggered star formation towards the entire IC 1396A region.

\subsubsection{Outer clusters (C-3, C-4, and C-5)} \label{c3_c4_c5}
The outer clusters (C-3, C-4, and C-5) differ from the inner clusters based on their astrometry properties (see Figure \ref{parallax_propermotion}). C-3 is linked with BRC IC 1396 N, C-4 with SFO 37, and C-5 in the northwest boundary of IC 1396. The mean age of C-3 is slightly lower than C-1 (refer Table \ref{stat_age_mass}). This indicates that the triggered star formation mechanism also forms the stars associated with IC 1396 N. The mean age of C-4 and C-5 appears slightly higher than all other clusters. In these two clusters, a significant fraction of stars of higher age is present.  
Earlier studies carried out by \citet{2008AJ....135.2323I}, and \citet{2014MNRAS.443.1614P} have already reported sequential star-formation in the direction of BRCs SFO 37 and SFO 39 (see Figure 1) due to the UV radiation impact of the exciting central star. The cluster C-4 is associated with SFO 39, but we do not detect any significant clustering towards SFO 37, as it is a small globule-like structure consisting of mainly a few embedded pre-main-sequence stars.

\subsection{Radial Velocity} \label{Rv_vel}
\begin{figure}
\centering
\includegraphics[scale=0.45]{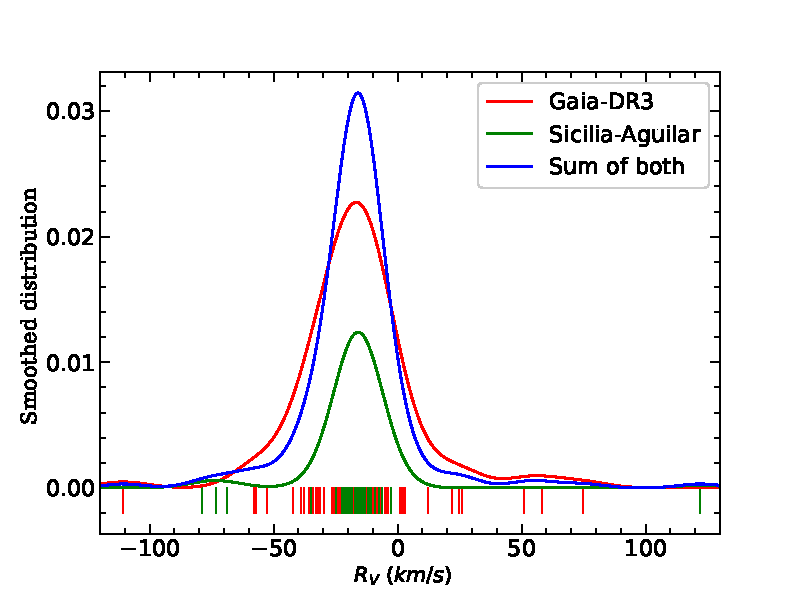}
\caption{Smoothed histogram distribution of $\rm R_V$ values for the 85 stars (red), 67 stars (green), and the total 152 stars (blue). We have $\rm R_V$ values for 85 stars from Gaia-DR3. For the 67 stars, $\rm R_V$ values taken from \citet{2006AJ....132.2135S}. To better represent the plot, the green curve is scaled down by $50\%$. The small vertical lines represent each star's $\rm R_V$ value. }
\label{hist_Rv}
\end{figure}

We searched for the stars with the radial velocity ($\rm R_V$) information in our member list. We obtained 107 stars with radial velocity information from Gaia-DR3. This is an improvement in $\rm R_V$ measurements in the Gaia-DR3 catalog compared to the DR2 catalog. Out of these 107 stars, 85 stars with good astrometry quality, i.e., $\rm RUWE<1.4$, are considered for further analysis.
The mean and median of $\rm R_V$ of the 85 stars are -16.30$\pm$1.28 and -16.56~$\rm km/s$, respectively.
To maximize the $\rm R_V$ measurements of the member stars of the complex, we also search for the $\rm R_V$ measurements in the literature. In previous work towards the region, \citet{2006AJ....132.2135S} has carried out high-resolution ($\rm R\sim 34000$) spectroscopic observations and obtained the radial velocity information for 136 stars. By cross-matching these stars with our Gaia-detected member lists, we find 78 stars in common, out of which 67 stars are of good astrometry quality, i.e., $\rm RUWE<1.4$. The mean and median of $\rm R_V$ of the 67 stars are -16.54$\pm$0.25 and -15.80~$\rm km/s$, respectively.
The $\rm R_V$ has a broad range for 85 stars compared to the list of 67 stars taken from \citet{2006AJ....132.2135S}. However, the mean and median values for both lists are similar. In Figure \ref{hist_Rv}, we display smooth histogram distribution for sources from both lists. In the figure, we scaled down the curve for the 67 stars by $50\%$ for a better representation. Smoothed distribution from this figure also suggests similar mean and median values found from the different lists. The spatial distribution of these 152 stars with $\rm R_V$ information is shown in Figure \ref{radec_pm}. Most of the stars are distributed within the central part of the complex, with few distributed all around the complex. Out of these 152 stars, 68 are members of the central cluster (C-1). We note that the properties of the complex and identification of different sub-groups in the complex by us are in close agreement with the recent work by \citet{2022arXiv221011930P}.

\section{Discussion} \label{diss}

\subsection{Kinematic properties of IC 1396} \label{kinematic}
\begin{figure}
\centering
\includegraphics[scale=0.37]{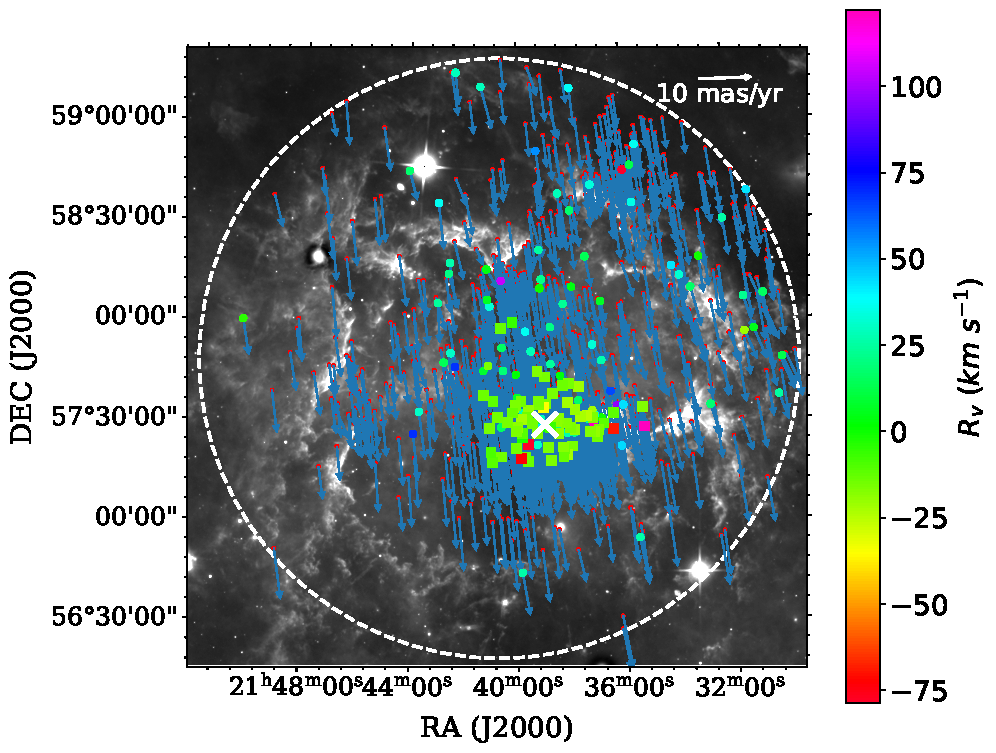}
\caption{Spatial distribution of the 1243 member stars as red dots on the WISE $\rm 22~\mu m$ band. The proper-motion values are shown as arrows. A reference arrow of 10~mas/yr is shown in the top right corner of the image. The 152 stars having $\rm R_V$ information are highlighted, where the 85 stars from Gaia based $\rm R_V$ are shown as solid circles and the 67 sources from \citet{2006AJ....132.2135S} are shown as square symbols. The colors of these objects mark their variation in $\rm R_V$, which is displayed as the color bar.}
\label{radec_pm}
\end{figure}

In Figure \ref{radec_pm}, we show the spatial distribution of the 1243 stars on the WISE $\rm 22~\mu m$ band as red dots, along with their proper-motion values as blue arrows. The magnitudes of $\rm \mu_{\alpha}cos\delta, and ~\mu_{\delta}$ gives the length of the arrow and the signs of $\rm \mu_{\alpha}cos\delta, and ~\mu_{\delta}$ determine the direction. All the arrows are scaled according to the white reference arrow of length 10~mas/yr. As seen from the plot, most stars are moving towards the south, one of the unique features observed towards the star-forming complex. In this section, we analyze the kinematics of the complex to shed more light on the internal motion of the member stars within the complex.

\subsubsection{Determination of 3-dimensional position and velocity} \label{3d_vel}
Since the complex IC 1396 is a relatively large star-forming complex, it is essential to inspect its physical structure and spatial distribution in Galactic cartesian coordinates, XYZ. We derive the XYZ coordinates for all the sources associated with IC 1396. The origin of the coordinate system is chosen to be Sun. In this system, the X-axis runs along the Sun-Galactic center with a positive direction toward the Galactic center, and the Y-axis is in the Galactic plane orthogonal to the X-axis with its positive direction along the Galactic rotation, the Z-axis is perpendicular to the Galactic plane, oriented in the direction of Galactic North Pole. Thus it makes a right-handed coordinate system. We used the Gaia-DR3 astrometric information of the detected stars and derived their 3-dimensional positions (X, Y, Z) and the heliocentric velocities (U, V, W). We have also computed the LSR velocities for each star along with the heliocentric velocities. The transformation of heliocentric to LSR velocity transformation made considering the solar motion velocities ($\rm U_0~=~11.1\pm0.7~km/s,~ V_0~=~12.2\pm0.47~km/s,~, and~ W_0~=~7.25\pm0.37~km/s$) from \citet{2010MNRAS.403.1829S}.

The majority of stars with $\rm R_V$ information lie towards the complex's central region. So to obtain the kinematic property, we focus only on the central cluster C-1. Table \ref{3dvel} lists the derived 3D-dimensional positions (X, Y, Z), the heliocentric velocities (U, V, W), and the LSR velocities of the 68 stars of the cluster C-1.

\begin{table*}
\tiny
\centering
\caption{3-dimensional position, heliocentric velocities, and LSR velocities of the 68 stars within the cluster C-1.}
\label{3dvel}
\begin{tabular}{cccccccccccccccc}
\\ \hline

Star No. & RA (2000) & DEC (2000) & X  & Y  & Z  & U    & V    & W    & u    & v    & w & $\rm \bf {\hat{r}_*~.~\bf{\delta v_*}}$ & \multicolumn{3}{c}{$\rm \bf {\hat{r}_*~\times~\bf{\delta v_*}}$}  \\
         & (degree)  & (degree)   &(pc)&(pc)&(pc)&(km/s)&(km/s)&(km/s)&(km/s)&(km/s)&(km/s) & (km/s)& \multicolumn{3}{c}{(km/s)} \\
\hline
1 & 325.1479 & 57.4754 &  -165.86 & 998.44 & 90.92 & 22.66 & -18.99 & -13.75 & 33.76 & -6.75 & -6.50 & -5.02 & -0.87 & -0.16 & 1.19\\ 
2 & 324.8110 & 57.3874 &  -150.43 & 925.01 & 87.06 & 17.96 & -6.24 & -12.17 & 29.06 & 6.00 & -4.92 & -7.60 & 3.56 & 1.14 & 0.50\\ 
3 & 324.9923 & 57.4759 &  -142.16 & 861.63 & 83.00 & 19.24 & -18.27 & -12.72 & 30.34 & -6.03 & -5.47 & 3.72 & -0.40 & 0.11 & 1.75\\ 
4 & 325.0083 & 57.5653 &  -170.96 & 1028.83 & 95.00 & 20.73 & -54.06 & -11.33 & 31.83 & -41.82 & -4.08 & -38.90 & 3.59 & 0.29 & -7.38\\ 
5 & 324.6100 & 57.4779 &  -135.03 & 832.29 & 83.10 & 16.58 & -14.38 & -11.84 & 27.68 & -2.14 & -4.59 & -0.66 & -0.90 & 0.10 & 3.83\\ 
6 & 324.5352 & 57.4465 &  -142.94 & 885.98 & 86.76 & 13.13 & -10.52 & -8.97 & 24.23 & 1.72 & -1.72 & -5.37 & -3.48 & -0.29 & 6.45\\ 
7 & 324.4657 & 57.4479 &  -143.82 & 894.13 & 87.71 & 18.56 & -13.26 & -10.34 & 29.66 & -1.02 & -3.09 & -1.56 & -2.31 & -0.42 & 1.64\\ 
8 & 324.7432 & 57.4737 &  -145.45 & 891.47 & 86.29 & 21.12 & -20.71 & -14.08 & 32.22 & -8.47 & -6.83 & 6.49 & 0.79 & 0.15 & 0.34\\ 
9 & 324.7699 & 57.4714 &  -137.55 & 842.19 & 82.85 & 21.43 & -30.57 & -13.51 & 32.53 & -18.33 & -6.26 & 16.20 & -0.46 & 0.05 & 1.60\\ 
10 & 324.8082 & 57.4760 &  -141.29 & 863.34 & 84.10 & 16.27 & 5.70 & -11.68 & 27.37 & 17.94 & -4.43 & -20.46 & 0.53 & 0.17 & 1.10\\ 
\hline
\end{tabular}
\\(This table is available in its entirety as online material. A portion is shown here for guidance regarding its form and content.)
\end{table*}

\subsubsection{Kinematic properties of the stars} \label{kine_stars}
\begin{figure*}
\centering
\includegraphics[scale=0.25]{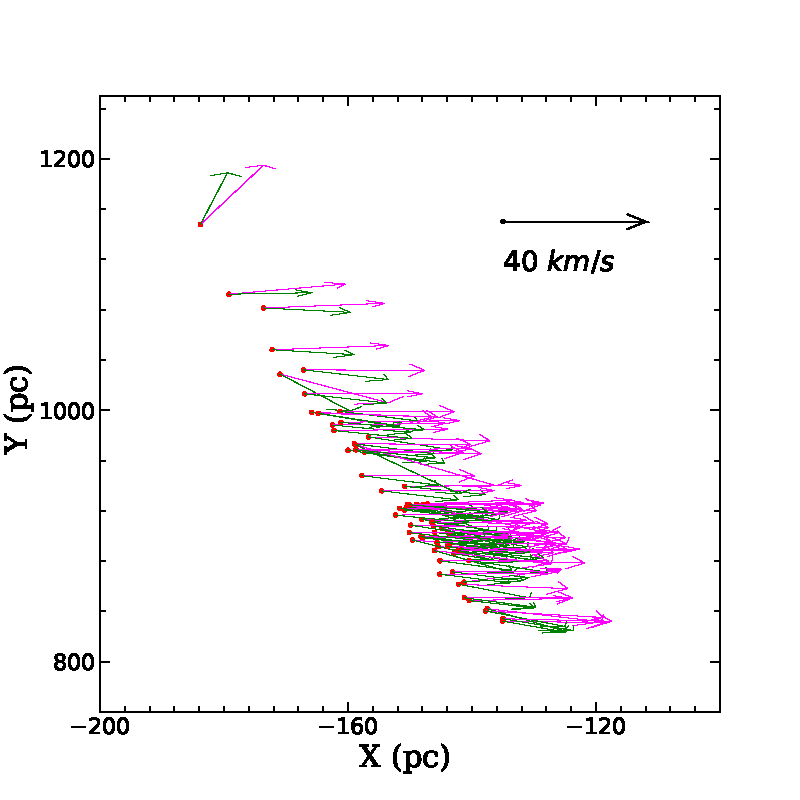}
\includegraphics[scale=0.25]{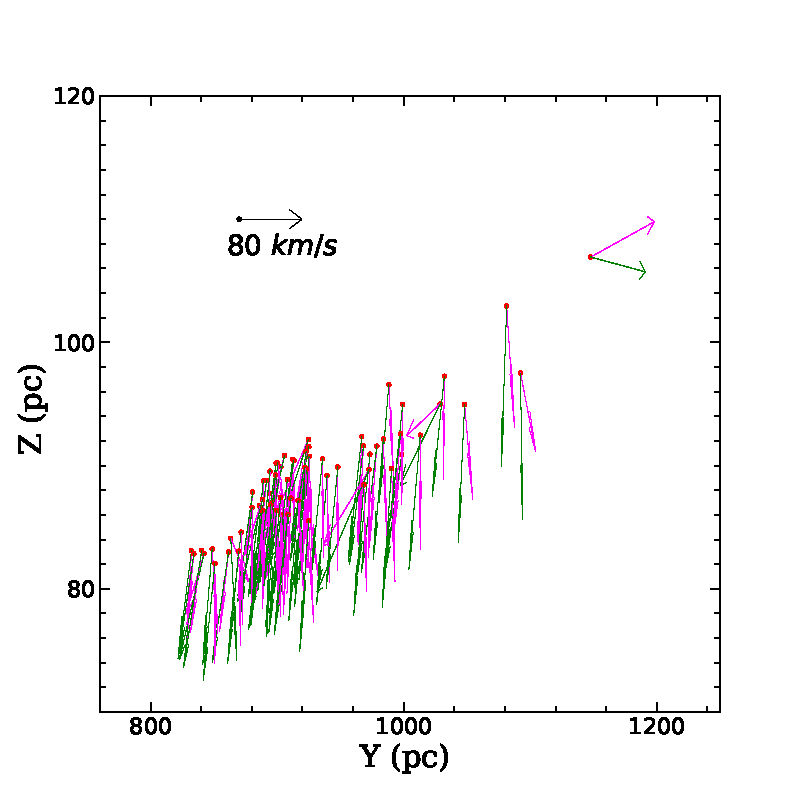}
\includegraphics[scale=0.25]{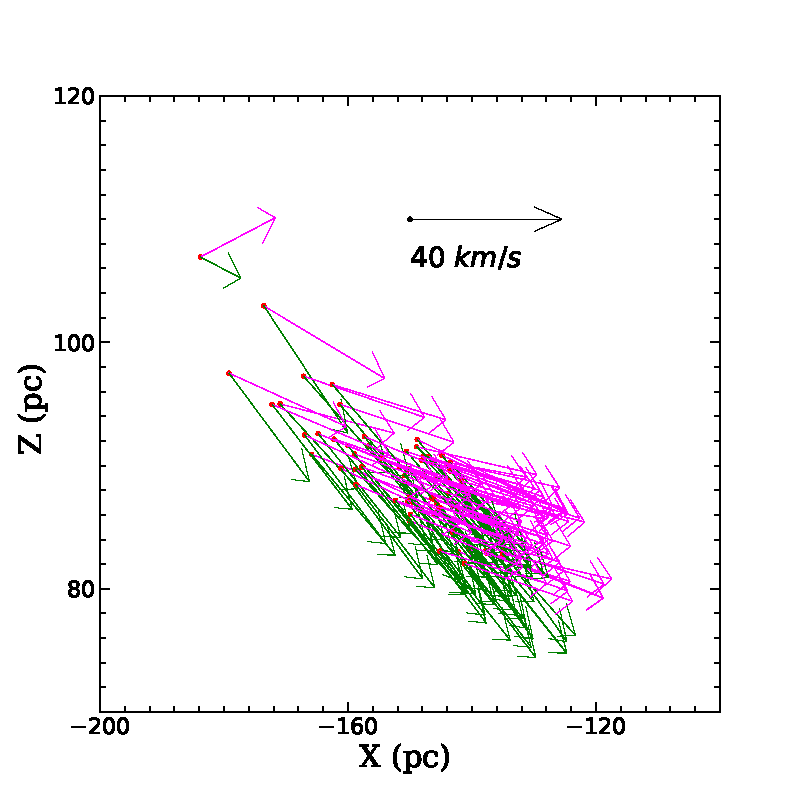}
\includegraphics[scale=0.25]{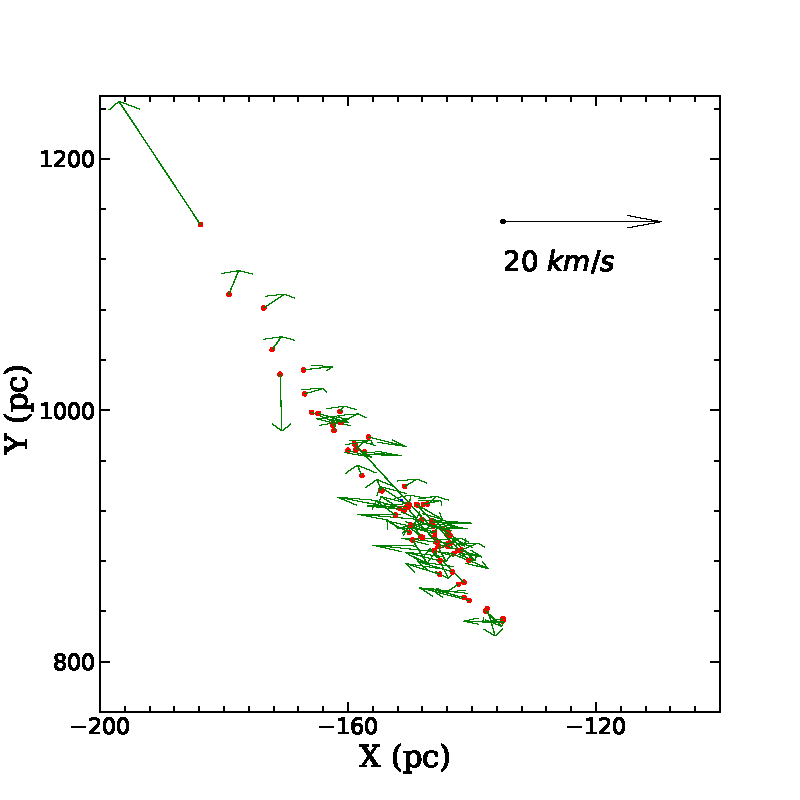}
\includegraphics[scale=0.25]{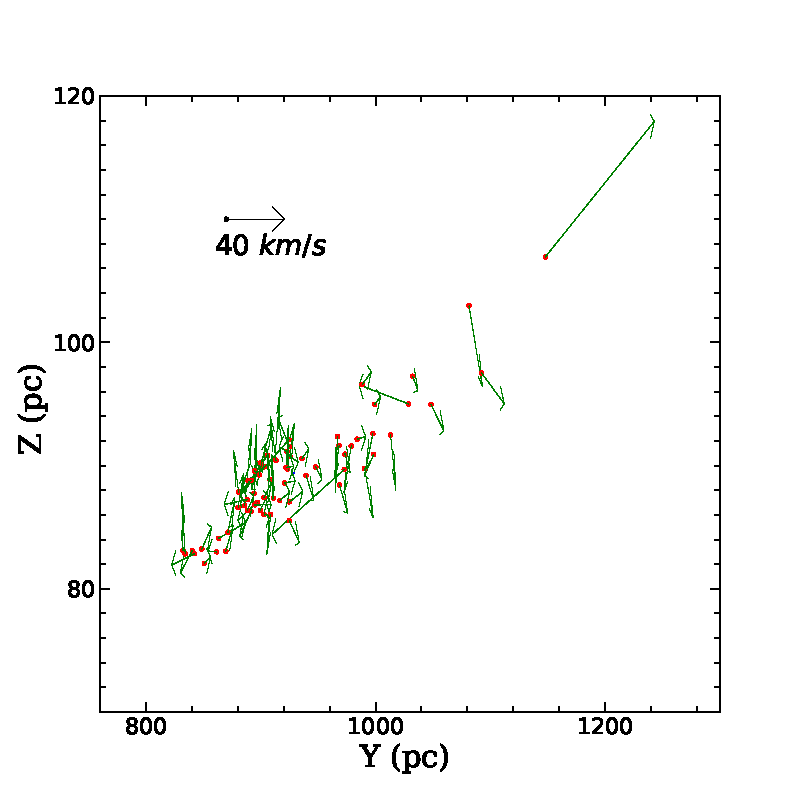}
\includegraphics[scale=0.25]{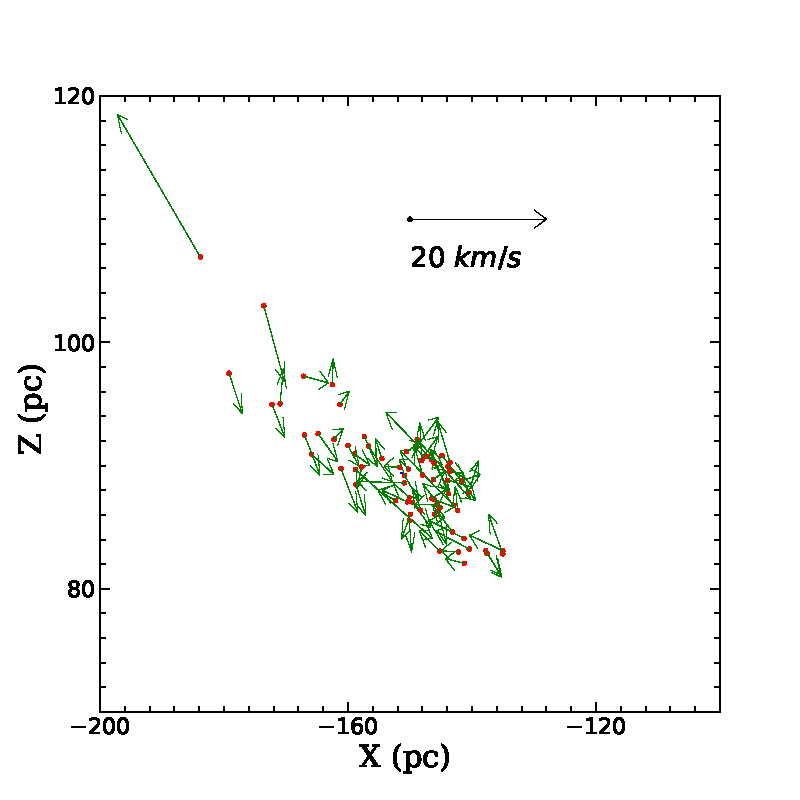}
\caption{Spatial distribution of the 68 stars of C-1 on XY, YZ, and XZ planes. Top: arrows represent the heliocentric (green) and LSR (magenta) velocities of the 68 stars, respectively. Bottom: Same as top, but the arrows represent the difference between the individual velocities and mean velocity. Details of the velocities are given in the text.}
\label{xyz_vel}
\end{figure*}

In Figure \ref{xyz_vel}, we show the spatial distribution of the 68 stars of C-1, which have radial velocity information in the XY, YZ, and XZ planes. In the top row, we display the heliocentric and LSR velocities. The heliocentric and the LSR velocities indicate the stars' bulk motion.

To investigate the stability of the cluster C-1, it is essential to analyze the internal kinematics of the stars. First, we derive the mean value of the velocities of the stars. The values are listed in Table \ref{3d_mean_vel}. To assess the internal motion of the stars, we calculate the difference in velocities $\rm (\delta u, \delta v, \delta w)$ of individual stars with respect to the mean value. In the bottom row of Figure \ref{xyz_vel}, we show the $\rm (\delta u, \delta v, \delta w)$. This displays the random movement of the stars with respect to the central velocity. 
This shows that $\rm (\delta u, \delta v, \delta w)$ of stars are canceling each other, and the mean values of $\rm (\delta u, \delta v, \delta w)$ are close to zero, indicating no real expansion.    
The three dimensional dispersion is $\rm \sigma = \sqrt{(\sigma u)^2 + (\sigma v)^2 + (\sigma w)^2}$ derived to be $\rm 16.56~km/s$.

Then we conduct a qualitative analysis of the relative motion of the stars within the complex in a similar manner carried out by \citet{2015ApJ...807..119R} for the Taurus complex. This analysis will provide an implication of the stability of the complex. Each star is located at a certain distance from the complex's center and moves with a relative velocity. We denote the separation from the complex center with a position vector $\rm \bf {r_*}$ and the relative velocity vector as $\rm \bf{\delta v_*}$. Each position vector is associated with a unit vector, which can be represented as $\rm \bf{\hat{r}_* = r_* / |r_*|}$, directing from the center of complex towards the location of each star. So the relative motion of stars with respect to the complex center can be used to analyze the two types of motions expansion or contraction and rotation. The expansion and contraction properties can be gauged by looking at the directions of the position vector and the relative velocity vector. For expansion, $\rm \bf{\delta v_*}$ will be parallel to $\rm \bf {r_*}$ and for contraction $\rm \bf{\delta v_*}$ will be anti-parallel to $\rm \bf {r}_*$. Hence for expansion, the dot product ($\rm \bf {\hat{r}_*~.~\bf{\delta v_*}}$) should be a large and positive number, and for contraction, it should be a large and negative number. In a similar analogy, the cross product ($\rm \bf {\hat{r}_*~\times~\bf{\delta v_*}}$) will be small for both expansion and contraction. 
In other way, the cross product ($\rm \bf {\hat{r}_*~\times~\bf{\delta v_*}}$) will be higher for large-scale rotation, and the dot product ($\rm \bf {\hat{r}_*~.~\bf{\delta v_*}}$) will be minimal.

In the following, we derive the dot and cross products and list them in Table \ref{3dvel}. Since in both the dot and cross product parameters, we use the unit position vector $\rm \bf{\hat{r}_*}$, the values of both the parameters have similar velocities. The mean values of the parameters can be expressed with the equations $\rm v_{exp} = \overline{ \bf {\hat{r}_*}~.~\bf{\delta v_*} }$ and
$\rm v_{rot} = \overline{ \bf {\hat{r}_*}~\times~\bf{\delta v_*} }$.

\begin{table*}
\tiny
\centering
\caption{Mean values of heliocentric and LSR velocities and dispersions derived from the 68 stars of cluster C-1. Values of expansion and rotational velocities are listed in the table.}
\label{3d_mean_vel}
\begin{tabular}{cccccccccccccccccc}
\\ \hline
Cluster & $\rm \bar{U}$ & $\rm \bar{V}$ & $\rm \bar{W}$ & $\rm \bar{u}$ & $\rm \bar{v}$ & $\rm \bar{w}$ & $\rm \sigma u$ & $\rm \sigma v$ & $\rm \sigma w$ & $\rm \sigma$ & $\rm v_{exp}$ & \multicolumn{3}{c}{$\rm v^a_{rot}$}     \\
         &(km/s)&(km/s)&(km/s)&(km/s)&(km/s)&(km/s) & (km/s)&(km/s)&(km/s)&(km/s)&(km/s)& \multicolumn{3}{c}{(km/s)} \\
\hline
C-1 & 20.47 & -14.33 & -12.75 & 32.57 & -2.09 & -5.50 & 3.04 & 16.15 & 2.03 & 16.56 & 1.11 & -0.06 & 0.07 & 1.02  \\
\hline
\end{tabular}
\\ $\rm ^a$ Columns 13, 14, and 15 list values of the three components of the rotation velocity.  
\end{table*}

We derive the expansion velocity, $\rm v_{exp}$, to be 1.11~km/s. The derived rotation velocities are listed in Table \ref{3d_mean_vel}. From the CO maps \citet{1995ApJ...447..721P} have obtained an expansion velocity of the whole complex to be 5~km/s. Their analysis suggests that the gas within the complex is pushed away to the outskirts by the central massive star resulting in an expansion of the system. A similar expanision velocity is also observed by \citet{2022arXiv221011930P}. Though cluster C-1 is expanding, but its expansion is slow compared to the whole complex. This could be because young stars dominate the central region, and cluster C-1 is expanding slowly due to higher density.

Nearby Galactic clusters are expanding with similar velocities as of cluster C-1, observed by \citet{2019ApJ...870...32K}. Their study over a set of 28 Galactic clusters using Gaia-DR2, reported a typical expansion velocity of $\rm \sim0.5~km/s$. Similarly, the study conducted by \citet{2021ApJ...912..162P} of 13 open clusters within a distance of 500~pc using Gaia-EDR3 reported many clusters to be super-virial and expanding in nature.

\subsection{Star-formation history in IC 1396} \label{over_star_form}
IC 1396 is one of the nearby star-forming complexes dominated by feedback-driven star formation activity (see Section \ref{source_detl}). The energetic stellar wind from the central massive star has cleared up most of the gas, resulting in a cavity of radius $\sim1.5^{\circ}$. The large cavity can be seen at infrared wavelengths with photodissociation regions (PDRs) associated with the boundary of the complex (see Figure \ref{wise}). This massive feedback effect also forms BRCs and fingertip structures within the complex \citep{1991ApJ...370..263S, 2005A&A...432..575F, 2012MNRAS.421.3206S}. Here, we discuss the overall star formation history of the complex.

The spatial distribution of the member sources (see Figure \ref{rf_wise}) and their association with the BRCs all indicate the ongoing feedback-driven star formation activity within the complex. The mean age of the sub-clusters (see Section ) suggests a multi-generation star formation activity within the complex. However, the sub-clusters formation in the complex might have happened through a hierarchical process. To assess this nature, we conduct a KS test on the age of the two major groups of stars (see Section \ref{cluster}). One group is from the inner clusters (C-1 and C-2), and the other is from the outer clusters (C-3, C-4, and C-5). The $p-$ score of the KS test comes out to be 0.00026. This low value of $p-$ score indicates that a majority number of stars from both groups might have formed over a similar time scale. 
The hierarchical star formation could be due to the fractal and turbulent nature of the ambient cloud, where star formation can occur simultaneously or near simultaneously at different locations of the clouds \citep{2003MNRAS.343..413B,2018MNRAS.481..688G,2022MNRAS.510.2097T}. However, one limitation of our analysis is that we have probed stars using optical 
measurements. Thus many sources embedded in the BRCs might be missing in our analysis; as a result, the estimated ages of the groups associated with BRCs are likely upper limits.

Kinematics and age analysis of the embedded members are needed to understand whether the groups associated with BRCs are formed through entirely hierarchical collapse processes or whether stellar feedback from the central cluster has helped induce star formation in these clouds. In favorable conditions, stellar feedback can enhance or accelerate star formation in pre-existing clouds where star formation is already underway. In this case, one may have both older as well as the young population of sources. Observations show that young clusters tend to show typical velocity dispersion of 2~km/s \citep{2019ApJ...870...32K}. Thus, older stars can move $\sim2$~pc in 2~Myr of time, so inferences such as age gradient and elongated morphology, which are signatures of induced star formation as we move from ionizing sources to the tip of the BRCs, can be erased, particularly, if we are dealing with smaller groups or number of stars. Thus compressive spectroscopic and kinematic analysis of member stars in both the optical and infrared bands would be highly desirable to shed more light in understanding the formation of different sub-groups in the complex.

\section{Summary} \label{summ}
We use the high-precision Gaia-DR3 astrometry and photometry data and apply the machine learning algorithms to carry out the membership analysis of the complex. Using the identified members in this work, we study various star-formation properties of this complex. In the following, we report our significant findings from this work.

\begin{enumerate}
\item Using the Gaia-DR3 astrometry and photometry data and applying the supervised RF technique of the machine learning algorithm, we identify this complex's 1243 high probable member populations. The identified member population is of high quality, with 95\% stars having a relative parallax error of less than 20\%. More than 99\% stars have RUWE less than 1.4 suggesting they are of high astrometry quality. Of the 1243 stars, 731 are entirely new members identified in this work. This has significantly enhanced the reliable member population list for IC 1396.

\item The mean values of the parameters RUWE, parallax, $\rm \mu_{\alpha}cos\delta$, and $\rm \mu_{\delta}$ are 1.12, $1.085\pm0.003$~mas, $-1.194\pm0.002$~mas/yr, and $-4.215\pm0.004$~mas/yr, respectively. The spatial distribution of the parallax, $\rm \mu_{\alpha}cos\delta$, and $\rm \mu_{\delta}$ suggests that the total population is broadly segregated into two groups. Our KS test shows that proper motion parameters are the most distinctive astrometric features, distinguishing the stars projected in the two sub-groups.

\item The spatial distribution of the stars reveals the associated clusters. We use the NN method to identify 6 clusters (\# C-1A, C-1B, C-2, C-3, C-4, and C-5) towards IC 1396. C-1A and C-1B are the subclusters of the central cluster C-1. We study the statistical properties of stars lying within the subclusters.

\item Using the G vs. G-RP CMD and parsec isochrones, we estimate the age and mass of individual stars. The mean age derived from all the 1243 stars to $1.5\pm1.6$~Myr, matching with the estimations from previous studies. Using the completeness limit of 19~mag in the G band and distance to be 917~pc, we derive the mass completeness limit for the complex to be $\rm \sim0.1~M_{\odot}$. Thus suggesting the complex is associated very low massive population.

\item Of the 1243 stars, 152 good quality stars ($\rm RUWE<1.4$) have $\rm R_V$ measurements, out of which 85 stars $\rm R_V$ information from Gaia-DR3 and the remaining 67 stars from a high-resolution spectroscopic study of \citet{2006AJ....132.2135S}. The mean and median values of $\rm R_V$ derived from the 152 stars are $-16.41\pm0.72$ and 15.80~km/s, respectively.

\item We carry out a 3D kinematic analysis to understand the internal motion of stars within the central cluster C-1. We use the $\rm R_V$ values and astrometric data of the 68 stars of the cluster. We derive the 3-D cartesian positional and velocities of each star. To study the stability of the cluster, we derive the expansion velocity, which is low compared to the previous value derived based on CO maps. The low value of the expansion velocity of the cluster suggests a slow expansion compared to the whole complex. The slow expansion might be due to the higher density of recently formed young stars.

\item Considering the spatial distribution, association with BRCs, and age of stars, we study the overall star formation within the complex. The variation in the age of the sub-clusters suggests an ongoing multi-generation star formation process in the complex. However, the sub-clusters of the complex might have formed through a hierarchical process.

\end{enumerate}

We thank the anonymous referee for a constructive review of the manuscript, which helped in improving the quality of the paper. SRD acknowledges support from Fondecyt Postdoctoral fellowship (project code 3220162). This work presents results from the European Space Agency (ESA) space mission Gaia. Gaia data are being processed by the Gaia Data Processing and Analysis Consortium (DPAC). Funding for the DPAC is provided by national institutions, in particular the institutions participating in the Gaia MultiLateral Agreement (MLA). The Gaia mission website is https://www.cosmos.esa.int/gaia. The Gaia archive website is https://archives.esac.esa.int/gaia. This publication makes use of data products from the Wide-field Infrared Survey Explorer, which is a joint project of the University of California, Los Angeles, and the Jet Propulsion Laboratory/California Institute of Technology, funded by the National Aeronautics and Space Administration. This research has made use of the SIMBAD database, operated at CDS, Strasbourg, France. This work made use of various packages of Python programming language.

\bibliography{refer}{}
\bibliographystyle{aasjournal}


\appendix 

\section{Gaussian Mixture Model} \label{gmm_det}
GMM works on the simple principle of identifying the normally distributed sub-populations from the overall population. This model assumes that the data points are generated from a mixture of a finite number of Gaussian distributions with unknown parameters. In the GMM method, each data point will be categorized into cluster members or non-members, depending on its membership score (probability). The mixture models do not require prior knowledge of classifying sub-populations. This allows the model to learn the sub-population in an automated way. Since there is no previous knowledge of the sub-population assignment, this mixture model constitutes unsupervised machine learning. This technique is widely used in various fields, including astrophysics \citep{2012MNRAS.424.2832L,2013MNRAS.434.2229I,2016MNRAS.462.3243Z,2017MNRAS.469.3374C,
2017AJ....153..249H,2018ApJ...855...14K,2018AJ....156..121G,2018ApJ...869....9G}. Below we briefly describe the working principle of the GMM method.

If there are m clusters present in n-dimensional parameter space, then the probability distribution $\rm P(x)$ of a data x is estimated as the weighted summation of all the m-Gaussian components.

\begin{equation}
\rm P(x) = \sum_{k=1}^{m} w_k~ P(x~|~\mu_k,~ \Sigma_k)
\end{equation}

where, $\rm \Sigma_k$ is the covariance matrix, and $\rm w_k$ is the mixture weight of the k-th Gaussian component, which satisfies the condition $\rm \sum_{k=1}^{m} w_k = 1$. The distribution of individual Gaussian cluster is 

\begin{equation}
\rm P(x~|~\mu_k,~ \Sigma_k) = \frac{exp \left[\frac{-1}{2}(x-\mu_k) \Sigma^{-1}_{k}(x-\mu_k)\right] }{(2\pi)^{n/2} \sqrt{|\Sigma_k|}}
\end{equation}

where, $\rm \mu_k$ and $\rm \Sigma_k$ are the mean vector and covariance matrix of the k-th Gaussian component. The $\rm |\Sigma_k|$ is the determinant of $\rm \Sigma_k$. 

In GMM, the parameters $\rm w_k, \mu_k, and~\Sigma_k$ are determined using the unsupervised machine-learning technique, known as the expectation-maximization (EM) algorithm \citep{Dempster77maximumlikelihood,press2007numerical}. 
The maximum likelihood of the data strictly increases with each subsequent iteration, which implies that it is guaranteed to approach a local maximum. This algorithm does not assume any prior knowledge about clustering structures. The EM algorithm starts with an initial guess for N data points and learns the GMM parameters from the data. This process involves a few steps, which is described in detail in \citet{2012MNRAS.424.2832L}. After calculating the distribution parameters, the distribution probability $\rm P(x~|~\mu_k,~ \Sigma_k)$ for each data point x can be estimated. 

Before carrying out the clustering analysis, it is essential to normalize the data. This data normalization is often required for similarity measures (e.g., Euclidian distance), which are sensitive to the differences in magnitudes or scales \citep{2018AJ....156..121G,2018ApJ...869....9G}. In our case, we have done the data normalization following the discussions made in \citet{2018AJ....156..121G}. If N stars have an n-dimensional parameter space, the normalized parameter in the jth dimension $\rm X^j_i$ is defined as:

\begin{equation}
\rm X^j_i = \frac{x^j_i - Med(x^j)}{\sigma x^j}
\end{equation} 

where, $\rm x^j_i$ the original parameter, $\rm Med(x^j)$ is the median of $\rm x^j$ distribution, and $\rm \sigma x^j$ is its standard deviation.

\section{Random forest classifier efficiency} \label{rf_eff}

\begin{figure}
\centering
\includegraphics[scale=0.3]{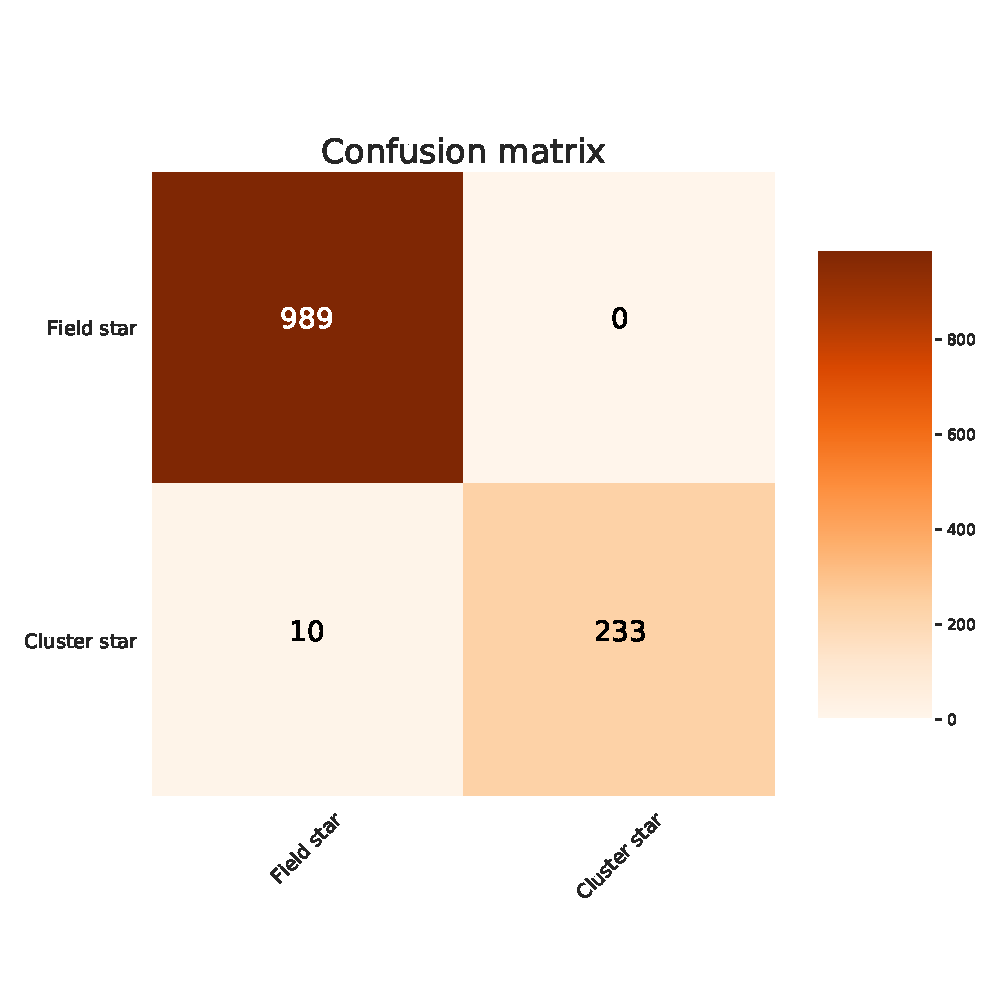}
\caption{Confusion matrix generated by RF method. The cluster stars and field stars are displayed in the plot. }
\label{rf_confusion}
\end{figure}

As explained in Section \ref{GMM}, within a small circular area of radius $30\arcmin$, we find 577 stars as probable members, and 2503 stars to be non-members. Using this result of GMM, we construct a reliable training set. This is quite important since the RF method is highly dependent on the training set. Since RF is handy in handling large dimensions, we use 11 input parameters in this work. The input parameters set include five position parameters: coordinates, proper motions, parallax, and six photometric parameters such as magnitudes in G-band, BP-band, RP-band, BP-RP color, and BP-G and G-RP color. Hence, we construct the RF classifier using the 11-dimensional reliable training set and test its accuracy. For this purpose, we use 60\% of the input 3080 stars to train the RF classifier and the remaining 40\% data to test the accuracy. So in our case, out of 3080 stars, 1848 stars are used to train the RF method, and the remaining 1232 stars are used to test how well the machine gets trained in recognizing the member stars and the field stars. The machine itself randomly performs the choice of training and test sets. We obtain a high accuracy of 0.99 while running the RF method over the test data set. The confusion matrix shown in Figure \ref{rf_confusion} presents the RF method's high accuracy. This confusion matrix shows how the machine identifies the sources based on training. As can be deduced from the confusion matrix, out of 1232 sources used to test the machine's accuracy, the machine successfully identified 989 non-member or field stars and 233 cluster member stars. The machine is confused with only a few field and cluster member stars during classification. This exercise demonstrates the effectiveness of the RF method.

Table \ref{rf_impt} provides the relative importance of 11 input parameters found by RF while providing the membership probability. We see that the proper motion in ra ($\rm \mu_{\alpha}cos\delta$) has maximum relative importance in membership identification compared to other parameters. The proper motion in dec ($\rm \mu_{\delta}$) also has relatively high importance in segregating member and non-member stars. However, in our case, the color terms (BP-G and BP-RP) and the magnitude (RP-mag) get higher importance in correctly identifying members and non-members. The reason is due to the filtering of stars using CMDs during the GMM method (see Section \ref{GMM}). Usually, proper motions play a dominant role in cluster identification; in our case, we also observe the same. The coordinates of the stars (RA, DEC) have minor importance in membership identification. In the previous analysis, \citet{2018AJ....156..121G,2018ApJ...869....9G} also obtain a similar result in the regions NGC 6405 and M67. It is worth mentioning here that while running RF, there is no need for data normalization as was done for the GMM method.

\begin{table}
\centering
\caption{Relative importance of the 11 input parameters}
\label{rf_impt}
\begin{tabular}{cc}
\\ \hline \hline
Parameter & Releative importance \\
\hline
RA &  0.010 \\
DEC & 0.009 \\
Parallax & 0.027 \\
$\rm \mu_{\alpha}cos\delta$ & 0.172 \\
$\rm \mu_{\delta}$ & 0.128 \\
G-mag & 0.102 \\
BP-mag & 0.051 \\
RP-mag & 0.138 \\
BP-RP & 0.140 \\
BP-G & 0.169 \\
G-RP & 0.052 \\
\hline
\end{tabular}
\end{table}

\section{CMD plots of stars with $\rm P_{RF}\leqslant0.5$} \label{cmd_nonmemb}
\begin{figure}
\centering
\includegraphics[scale=0.35]{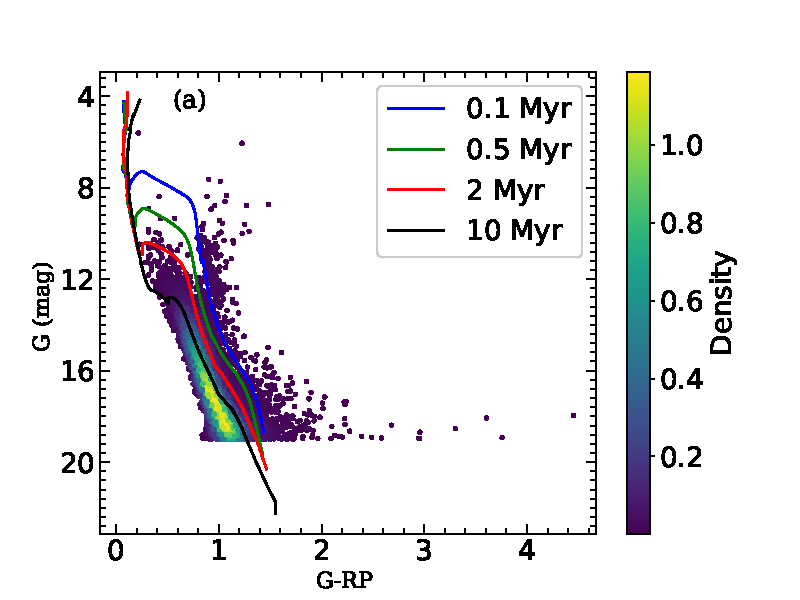}
\includegraphics[scale=0.35]{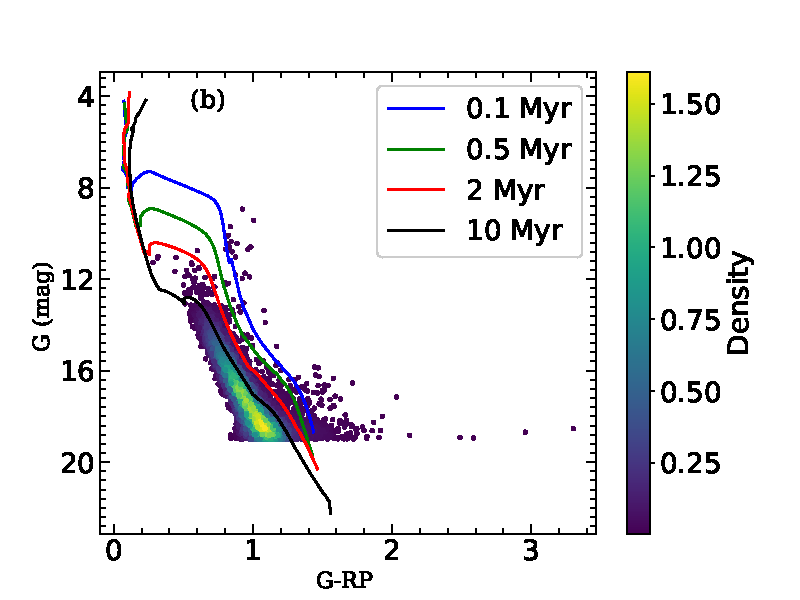}
\includegraphics[scale=0.35]{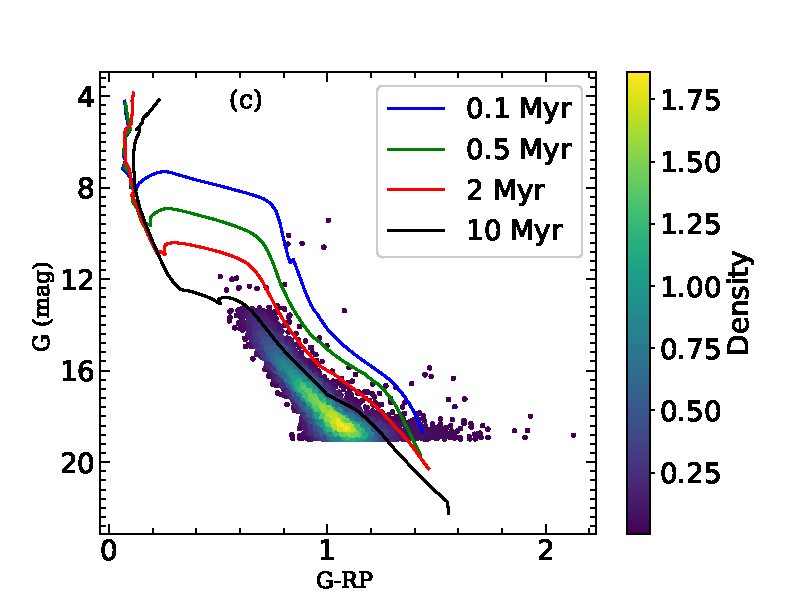}
\includegraphics[scale=0.35]{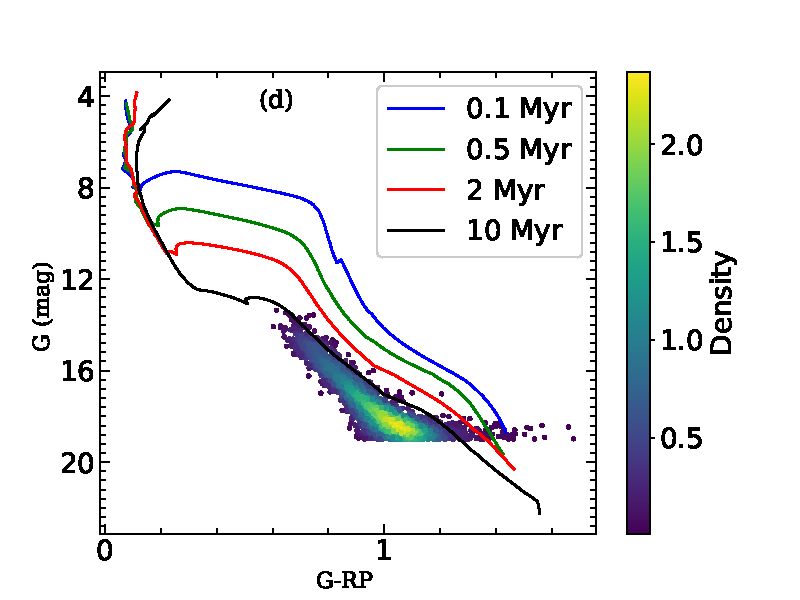}
\caption{Same as Figure \ref{g_rp_cm_iso}, but for the stars with $\rm P_{RF}\leqslant50\%$ and G band less than 19~mag. (a) All stars and (b), (c), and (d) for stars with $\rm P_{RF}\leqslant10\%,~ 5\%,~ and~ 1\%$, respectively. Colour-bar displays the variation in the density distribution of sources.}
\label{g_rp_cmd_non}
\end{figure}

Generally, stars with $\rm P_{RF}<50\%$ are non-member stars. Here in Figure \ref{g_rp_cmd_non}, we show the CMDs of stars retrieved with $\rm P_{RF}<50\%$ and G band less than 19~mag. Stars laying within different $\rm P_{RF}$ values are shown here. This is to check their location on the CMD. Out of the total stars with $\rm P_{RF}<50\%$ and G~mag less than 19, the majority ($70\%$) stars lie within $\rm P_{RF}<10\%$. The stars with $\rm P_{RF}<1\%$ are the most likely non-member. However, stars with higher probability spread on the plot. This discussion aims to shed light on the nature of stars with different probabilities. This is to stress the fact that the member and non-member stars should be chosen carefully in this type of membership analysis, where the magnitude and color terms will play a major role in segregating members and non-member stars. An overlap in the magnitude and color terms of both member and non-member stars will lead to the failure of effective training of the machine.

\end{document}